\documentclass[12pt,a4paper]{article}
\usepackage{amsmath,amssymb,extarrows,mathtools,graphicx,subfigure,setspace}
\usepackage[all]{xy}
\usepackage{epsfig,comment,geometry,comment,color,cite,slashed}
\usepackage[active]{srcltx}
\usepackage[affil-it]{authblk}
\usepackage{fullpage}
\numberwithin{equation}{section}
\newcommand{\be}{\begin{equation}}
\newcommand{\ee}{\end{equation}}
\newcommand{\bea}{\begin{eqnarray}}
\newcommand{\eea}{\end{eqnarray}}
\newcommand{\nn}{\nonumber}

\newcommand{\kk}{\mathbf{k}}
\newcommand{\vv}{\mathbf{v}}

\newcommand{\pp}{\mathbf{p}}
\newcommand{\eps}{\varepsilon^{a}\vspace{0.01cm}_{b c}}
\newcommand{\pv}{\partial_{v}}
\newcommand{\pu}{\partial_{u}}

\begin{document}
\begin{titlepage}
\vspace{10mm}
\begin{flushright}
 %...\\
\end{flushright}

\vspace*{20mm}
\begin{center}
{\Large {\bf Holographic Renormalization of  3D \\ Minimal Massive Gravity   }\\
}

\vspace*{15mm}
\vspace*{1mm}
{Mohsen Alishahiha${}^a$, 	Mohammad M. Qaemmaqami${}^b$, Ali Naseh${}^b$ and
Ahmad Shirzad${}^{c,b}$ }

 \vspace*{1cm}

{\it ${}^a$ School of Physics, Institute for Research in Fundamental Sciences (IPM)\\
P.O. Box 19395-5531, Tehran, Iran \\  
%${}^b$ Department of Physics, Sharif University of Technology,\\
%P.O. Box 11365-9161, Tehran, Iran\\
${}^b $ School of Particles and Accelerators\\Institute for Research in Fundamental Sciences (IPM)\\
P.O. Box 19395-5531, Tehran, Iran \\ 
${}^c $ Department of Physics, Isfahan University of Technology\\
P.O.Box 84156-83111, Isfahan, IRAN
}

 \vspace*{0.5cm}
{E-mails: {\tt \{alishah, m.qaemmaqami, naseh, shirzad\} @ipm.ir, }}%

\vspace*{1cm}
%%\maketitle
\end{center}

\begin{abstract}

We study holographic renormalization of 3D minimal massive gravity using 
the Chern-Simons-like formulation of the model. We explicitly present Gibbons- Hawking term as well as all counterterms needed 
to make the action finite in  terms of dreibein  and spin-connection. This can be used to find 
correlation functions of stress tensor of holographic dual  field theory. 
\end{abstract}

\end{titlepage}

%%%%%%%%%%%%%%%%%%%%%%%%%%%%%%%%%%%%%%%%%%%%%%%%%%%%%%%%%%%
\setcounter{footnote}{0}
\addtocontents{toc}{\protect\setcounter{tocdepth}{4}}
\setcounter{secnumdepth}{4}
%\tableofcontents
%%%%%%%%%%%%%%%%%%%%%%%%%%%%%%%%%%%%%%%%%%%%%%%%%%%%%%%%%%%%
\section{Introduction}
 %%%%%%%%%%%%%%%%%%%%%%%%%%%%%%%%%%%%%%%%%%%%%%%%%%%%%%%%%%%%
 
Topologically massive gravity (TMG) is a three dimensional gravity whose action consists 
of Einstein gravity with a cosmological constant  plus the gravitational Chern-Simon 
term \cite{{Deser:1982vy},{Deser:1981wh}}. The model has two free parameters (cosmological 
constant and the coefficient of Chern-Simon term) and for generic values of the parameters the 
corresponding equations of motion admit several solutions  including 
AdS, BTZ  and warped AdS black holes  solutions\cite{{Moussa:2003fc},{Bouchareb:2007yx},
{Anninos:2008fx},{Moussa:2008sj}}.  It is, however, known that TMG suffers from the fact that the 
energy of graviton and the mass of BTZ black holes cannot be positive at the same time
(see {\it e.g.} \cite{Li:2008dq}).

In order to circumvent the above problem, the authors of  \cite{Bergshoeff:2014pca}, proposed  
a new model, named Minimal Massive Gravity (MMG), which is a consistent ghost free 
and non-tachyonic three dimensional theory. 
Indeed the proposed model which is a natural extension of TMG has one more tunable 
parameter, though the same as TMG, it has a single massive mode above a
flat space or an  AdS vacuum.

A peculiar feature of MMG is that it does not have an action in the metric formulation, though 
it is possible to wire an action in terms of dreibein.  More precisely the Lagrangian 3-form of MMG is  
 \cite{Bergshoeff:2014pca} 
\bea\label{L.MMG}
L(e,\omega, h) &=& -\sigma e\cdot R(\omega) +\frac{\Lambda_{0}}
{6} e\cdot e\times e + h\cdot T(\omega)+\frac{1}{2\mu}
(\omega\cdot d\omega +\frac{1}{3} \omega\cdot \omega\times 
\omega) \cr &&+\frac{\alpha}{2} e\cdot h\times h,
\eea
where $\sigma$ is a sign, $\Lambda_0$ is a cosmological constant, $e$ and $\omega$ are  dreibein  
and dualised spin-connection, respectively.  $\mu$ is a  mass parameter of the model.
 Moreover in terms of these variables 
the Lorentz covariant torsion and curvature 2-forms  are given by 
\bea
T(\omega) = d e + \omega\times e,~~~~~~ R(\omega) 
= d\omega +\frac{1}{2} \omega \times \omega.
\eea
Here $h$ is an extra field that for $\alpha=0$ case it may be  thought of as a  Lagrange multiplier to 
impose the zero torsion  constraint. Indeed in this case the model reduces to the TMG model. 

To explore different properties of the model, it is useful to obtain   equations of motion derived 
from the action \eqref{L.MMG}. It is,  however, important to note  that in order to find the equations 
of motion one should  make sure that the boundary terms appearing due to the variation of the 
action could consistently  be removed from the variation.  This may be done
by  imposing proper boundary conditions or/and adding  certain  boundary terms.
It is indeed the aim of this paper to explore this point for MMG model using the 
holographic renormalization method\cite{Henningson:1998}. 

 We note, however,  that since the model does not 
have an action in the metric formulation one needs to study  holographic renormalization of the
MMG model in terms of its Chern-Simons like formulation.  Holographic renormalization for 3D 
Einstein gravity and 
3D topologically massive gravity in driebein formalism  have been studied in \cite{Banados:
1998ys,Miskovic:2006tm} and \cite{Blagojevic:2013bu,Grumiller:2015xaa}, respectively\footnote
{See \cite{{Alishahiha:2010bw}, {Loran:2013fca}} for holographic renormalization of other 3D gravities}. 

Different aspects of MMG model have been studied in {\it e.g.}
\cite{Afshar:2014ffa,Arvanitakis:2014yja,Tekin:2014jna ,Alishahiha:2014dma, 
Arvanitakis:2014xna,Giribet:2014wla,Setare:2014zea
,Arvanitakis:2015yya,Altas:2015dfa,Yekta:2015gja,Deger:2015wpa}.
In particular it was shown that the model has a critical point at which it exhibits 
logarithmic solutions. The corresponding solution may provide  gravitational 
descriptions for  dual logarithmic conformal field theories. Using the procedure developed 
\cite{Grumiller:2009mw} two point functions of stress energy and its logarithmic partner
have been obtained in  \cite{Alishahiha:2014dma} where the new anomaly has also been 
read from the expressions of the corresponding  two point functions. Here we will re-derive 
these results rather 
rigorously using the holographic renormalization method. To do so, we will have to carefully 
study  variational principle of the model and moreover to find any possible counterterms 
necessitate to make the model finite.

The paper is organized as follows. In the next section we will consider asymptotic analysis 
of the equations of motion of MMG model in the first order formalism. This study can be used to 
examine the validity of the variational principle of the model which is considered in section three.
In section four we will study the on-shell action of the model where we will obtain the necessary 
counterterms to make the  action  finite. Using the resultant finite on-shell action at the critical point
 we will obtain two point functions of holographic stress tensor and its logarithmic partner 
 in section five. The last section is devoted to conclusions.

\section{Asymptotic analysis}

In this section we would like to further  study  asymptotic behavior of the linearized  
equations of motion of MMG model.  The linearized equations of motion of the MMG model
above an AdS vacuum in the first order formalism have been considered 
in \cite{Bergshoeff:2014pca}. In what follows we will first review  the the relevant part of 
the paper  \cite{Bergshoeff:2014pca}
and then  we will solve the corresponding equations  asymptotically in the 
 Fefferman-Graham gauge. 
 
 To proceed we  
note that although for generic $\alpha\neq 0$  the torsion $T(\omega)$ is non-zero, one may define 
a new torsion free spin connection $\Omega=\omega+\alpha h$ by which the Lagrangian 
3-form \eqref{L.MMG}  reads  \cite{Bergshoeff:2014pca} 
\bea\label{Modified.L.MMG}
L(e,\Omega, h) &=& -\sigma e\cdot R(\Omega) +\frac{\Lambda_{0}}{6} 
e\cdot e\times e + h\cdot T(\Omega) +\frac{1}{2\mu}
(\Omega\cdot d \Omega+\frac{1}{3}\Omega\cdot \Omega \times \Omega)\cr
 &&
-\frac{1}{2\mu} 
\bigg(\alpha \Omega\cdot Dh+\alpha h\cdot R(\Omega)
-\alpha^2 h\cdot Dh+\frac{\alpha^3}{3}h\cdot h\times h\bigg)
\cr &&+\sigma\alpha e\cdot Dh
-\frac{\alpha}{2}(1+\sigma\alpha)e\cdot h\times h.
\eea
In this notation, assuming to have a well defined variation principle, the corresponding  
equations of motion are 
\bea  \label{e.o.m}
&& E_{e} = -\sigma R(\Omega)+(1+\sigma\alpha)
Dh+\frac{\Lambda_{0}}{2}
e\times e-\frac{\alpha}{2}(1+\sigma\alpha)h\times h=0, \\
&& E_{\Omega} = -\sigma T(\Omega)+(1+\sigma\alpha)e\times h+
\frac{1}{2\mu}\bigg( 2R(\Omega)+\alpha^{2}h\times h-
2\alpha Dh\bigg)=0,\cr 
&& E_{h} = (1+\sigma\alpha)T(\Omega)-\alpha(1+\sigma\alpha)
e\times h+\frac{1}{2\mu}\bigg(2\alpha^2 Dh-2\alpha 
R(\Omega)-\alpha^{3}h\times h\bigg)=0.\nn 
\eea
Here the covariant derivative is defined by  $DA = dA + \Omega \times A$.
Using the equations of motion for $\Omega$ and $h$ one can see that  $\Omega$ is 
a torsion free spin-connection,
\be\label{Torsion}
T(\Omega) =0,
\ee
by which the other equations of motion may be recast into the following forms
\bea\label{Two.e.o.m}
&& \sigma R(\Omega) -(1+\sigma\alpha) Dh 
-\frac{\Lambda_{0}}{2}
e\times e +\frac{\alpha}{2}(1+\sigma\alpha)h\times h =0, \cr 
&& R(\Omega)-\alpha Dh+\mu(1+\sigma\alpha)e\times h
+\frac{\alpha^{2}}{2}h\times h=0.
\eea

These  equations of motion 
exhibits an AdS vacuum solution with radius $l$ which can be given in terms of the 
parameters of the action as follows
\be\label{AdS.parameters}
\frac{\Lambda_{0}}{\mu^{2}}
 = -\frac{\sigma}{\mu^{2} l^{2}} 
  +\alpha(1+\sigma\alpha)C^{2},\;\;\;\;\;\;\;\;\;\;{\rm with}\;\;C = -\frac{(\alpha\Lambda_{0}l^{2}-1)}{2\mu^{2}l^{2}(1+\sigma\alpha)^{2}}.
\ee
Note that for this solution one also  gets $h = C\mu e$. 
It is then natural to study small fluctuations above 
this solution. Denoting the  vacuum solution by $\bar{e}, \bar{\Omega}, \bar{h}$, then 
a general perturbation may be written as 
\bea\label{perturbation}
e =\bar{e}+\kk,~~~~~~\Omega=\bar{\Omega}+\vv,~~~~h=C\mu(\bar{e}+\kk)+\pp,
\eea
where $\kk, \vv$ and $\pp$ are small perturbations of ${e},{\Omega}$ and $h$ respectively. 
Plugging  this anstatz into the equations of motion given by  (\ref{Torsion}) 
and (\ref{Two.e.o.m}) and using  (\ref{AdS.parameters}) one arrives at
\be\label{Linearised.e.o.m}
 \bar{D}\kk +\bar{e}\times \vv =0, \,\,\,\,\,
\bar{D}\vv+\bar{e}\times [\mu(1+\sigma\alpha)^{2} \pp +\frac{1}{l^{2}} \kk ]=0, \,\,\,\,\,
\bar{D}\pp +M \bar{e}\times\pp =0,
\ee
where $M = \mu(\sigma(1+\sigma\alpha)-\alpha C)$. 

Now the aim is to solve these linearized equations. To do so we will proceed as follows. 
By making use of the first equation in \eqref{Linearised.e.o.m} one can find $\vv$ in terms of 
$\kk$. More precisely one has
\bea\label{v}
\vv_{\mu}^{a} =-(\bar{e})^{-1}\epsilon^{\lambda\rho\nu}
\big(\bar{e}_{\mu b}\bar{e}_{\lambda}^{a}-\frac{1}{2}
\bar{e}_{\mu}^{a}\bar{e}_{\lambda b}\big)
\bar{D}_{\rho}\kk_{\nu}^{b}.
\eea
Note that this expression is exactly the same as that for 3D Einstein gravity\cite{Merbis:2014vja}.
This is due to the fact that the first  equation in \eqref{Linearised.e.o.m} is, indeed,
the torsionless condition appearing in both models.

Solving the second equation of \eqref{Linearised.e.o.m} for $\pp$ gives
\bea\label{p}
\pp_{\mu}^{a} = -\frac{1}{\mu(1+\sigma\alpha)^{2}}
\left((\bar{e})^{-1}\epsilon^{\lambda\rho\nu}
(\bar{e}_{\mu b}\bar{e}_{\lambda}^{a}-\frac{1}{2}\bar{e}_
{\mu}^{a}\bar{e}_{\lambda b})
\bar{D}_{\rho}\vv_{\nu}^{b}+\frac{1}{l^{2}}\kk_{\mu}^{a}\right),
\eea
where $\vv$ is given by (\ref{v}). Finally utilizing the expressions of $\vv$ and $\pp$ and from the last equation of 
\eqref{Linearised.e.o.m} one arrives at 
\bea\label{Lin.MMG}
&&\epsilon^{\lambda\mu\nu}\epsilon^{\gamma
\rho\sigma}\epsilon^{\eta\alpha\beta}
(\bar{e}_{\nu b}\bar{e}_{\gamma}^{a}-\frac{1}{2}
\bar{e}_{\nu}^{a}\bar{e}_{\gamma b})
(\bar{e}_{\sigma c}\bar{e}_{\eta}^{b}-\frac{1}{2}
\bar{e}_{\sigma}^{b}\bar{e}_{\eta c})
\bar{D}_{\mu}\bar{D}_{\rho}\bar{D}_{\alpha}\kk_{\beta}^{c}+ \cr
&&+M \epsilon^{\lambda\mu\nu}\epsilon^{\gamma
\rho\sigma}\epsilon^{\eta\alpha\beta}
\eps\bar{e}_{\mu}^{b}(\bar{e}_{\nu f}\bar{e}_{\gamma}^{c}
-\frac{1}{2}\bar{e}_{\nu}^{c}\bar{e}_{\gamma f})
(\bar{e}_{\sigma d}\bar{e}_{\eta}^{f}
-\frac{1}{2}\bar{e}_{\sigma}^{f}\bar{e}_{\eta d})
\bar{D}_{\rho}\bar{D}_{\alpha}\kk_{\beta}^{d}\cr
&&-\frac{\bar{e}^{2}}{l^{2}}\epsilon^{\lambda\mu\nu}
\bar{D}_{\mu}\kk_{\nu}^{a}
-\frac{\bar{e}^{2}}{l^{2}}M \epsilon^{\lambda\mu\nu}\eps
\bar{e}_{\mu}^{b}\kk_{\nu}^{c} = 0,
\eea
that is a  third order differential equation for  $\kk$.
 This equation should  be compared with the third order differential equation 
 obtained in \cite{Bergshoeff:2014pca} for $\kk_{\mu\nu}=\tilde{e}^a_\mu \kk_{\nu a}$
 where it was shown that the corresponding equation can be factorized into three first 
 order differential equations. Of course since we are interested in the holographic renormalization 
 of the model in the dreibein formulation in what follows we will directly work with $\kk_\mu^a$
 fields. Moreover in the context of holographic renormalization it is more appropriate to work in the 
 Fefferman-Graham coordinates. 

Actually an asymptotically locally  AdS geometry in the Fefferman-Graham (FG) coordinates  may be 
given  as follows\footnote{For  simplicity from now on we set  $l=1$.}
\be\label{FG}
ds^{2} = \frac{dr^{2}}{4r^{2}}+\frac{1}{r}g_{i j}(r,x^{\alpha})dx^{i}dx^{j}.
\ee
It can be verified that a dreibein which could reproduce  the above metric in the FG gauge has the 
following structure\cite{Blagojevic:2013bu}
\be\label{FG.dri}
e_{r}^{a} = (\frac{1}{2r},0,0),~~~~~e_{i}^{a} =(0,e_{i}^{2},e_{i}^{3}).
\ee 
Here we are using a notation in which  $x^{\mu}=(r,x^{i})$ 
and $x^{i}\equiv (u,v)$. For the vacuum  $AdS_{3}$ solution 
 with $\bar{g}_{uu}=\bar{g}_{vv} =0, \bar{g}_{uv} =1/2$, one has
\bea\label{drie.FG.AdS}
\bar{e}_{\mu}^{a} = \left(
  \begin{array}{ccc}
   \frac{1}{2r} & 0 & 0 \\
   0 & 0 & \frac{1}{2r} \\
   0 & 1 & 0 \\
  \end{array}
\right).
\eea
On the other hand for  perturbations in  spatial 
directions, $g_{ij} = \eta_{ij}+h_{ij}$, the corresponding  perturbations of dreibein 
respecting the  FG gauge are given by
\bea\label{k}
\kk_{r}^{a} = (0,0,0),~~~~~\kk_{i}^{a} =(0,\kk_{i}^{2},\kk_{i}^{3}).
\eea
Having expressed the FG gauge in the dreibein formalism it is important to make sure that
it has the right number of independent components. Naively from the above expression it seems
that  $\kk_{i}^{a}$ has four independent components. We note, however, that from 
the  integrability conditions of the  equations \eqref{Linearised.e.o.m}  as well as the explicit form 
of the equation \eqref{Lin.MMG} one can show that both
 $\pp_{\mu\nu}=\bar{e}_{\mu}^{a}\pp_{\nu a}$
and $\kk_{\mu\nu}=\bar{e}_{\mu}^{a}\kk_{\nu a}$ are symmetric tensors indicating that 
the actual number of independent components are three. 

More precisely from the linearized equations of motion (\ref{Linearised.e.o.m}) and 
using the symmetries of the background AdS${}_3$ geometry and integrability conditions
one gets 
\bea
\bar{e}^{a}\bar{e}\cdot \pp =0,
\eea  
showing that  $\pp_{\mu\nu}$ generally is  a symmetric tensor.

On the other hand taking  into account that the equation (\ref{Lin.MMG}) may be thought of 
as a vanishing tensor with  indices $\mathbf{E}^{\lambda a}$, one may find 
a new tensor by acting $ \bar{e}^{\eta}_{a} $ on it, {\it i.e.} $\mathbf{E}^{\eta\lambda}=
\bar{e}^{\eta}_{a} \mathbf{E}^{\lambda a}$. It is then easy to see that the equations of motion 
in the metric formulation is actually given by  $\frac{1}{2}(\mathbf{E}^{\eta\lambda}
+\mathbf{E}^{\lambda\eta})$,  on which only the symmetric combination of  
$\frac{1}{2}(\bar{e}^{a}_{\mu}\kk_{\nu a}+
\bar{e}^{a}_{\nu}\kk_{\mu a})$ appears in the equation. 
Therefore we could conclude that only symmetric part of $\kk_{\mu\nu}$ appears as 
a dynamical variable  which in turn indicates that the actual number of independent components of
$\kk^a_i$ is three\footnote{{We would like to thank 
Wout Merbis for a discussion on this point.}}. These three independent components may be chosen as 
$\kk_{u}^{2},\kk_{v}^{2},\kk_{v}^{3}$. In this notation one has $\kk^3_u=\frac{1}{2r}\kk^2_v$.

Using this gauge and  contracting the equation  \eqref{Lin.MMG} by  
$\bar{e}_{\lambda a}$  one finds \bea\label{TC}
-\frac{M}{\mu(1+\sigma\alpha)^{2}r}
\left(\frac{1}{2}\pv^{2}\kk_{u}^{2}-\pu\pv \kk_{v}^{2}+\kk_{v}^{2 
\prime} +r\pu^{2}\kk_{v}^{3}-2r
\kk_{v}^{2 \prime\prime}\right)=0,
\eea
where by ``prime'' we denote 
a  derivative with respect to radial coordinate $r$. In fact writing
down different components of (\ref{Lin.MMG}) in terms of $\kk_{i}^{a}$ gives the following
equations up to an overall factor of  $\frac{1}{\mu(1+\sigma\alpha)^{2}}$,
\bea\label{Lin.MMG.Final}
&&-\frac{1}{2}
\pv^{2}\kk_{u}^{2 \prime}+
\pu^{2}\kk_{v}^{3}+M\kk_{v}^{2 \prime\prime}
+r\pu^{2}\kk_{v}^{3 \prime} =0,\nn \\
&&2r
\pv \kk_{u}^{2 \prime\prime}-\pu\kk_{v}^{2 \prime}
+2r\pu\kk_{v}^{2 \prime\prime}+\pv\kk_{u}^{2 \prime}
-M\pv\kk_{u}^{2 \prime}+M\pu \kk_{v}^{2 \prime}=0 \nn \\
&&-4r^{2}
\pu\kk_{v}^{3 \prime\prime} +\pv\kk_{v}^{2 \prime}-2r\pv\kk_{v}^{2 \prime\prime}
-2Mr\pu\kk_{v}^{3 \prime}-10r\pu\kk_{v}^{3 \prime}+M\pv\kk_{v}^{2 \prime}
-2(M+1)\pu\kk_{v}^{3}=0 \nn \\
&&2r^{2}\pu^{2}
\kk_{v}^{3 \prime}-r\pv^{2}\kk_{u}^{2 \prime}+2r\pu^{2}
\kk_{v}^{3}+2Mr\kk_{v}^{2 \prime\prime}=0, \nn \\
&&-4r^{2}\kk_{v}^{3 \prime
\prime\prime} +\pv^{2}\kk_{v}^{2 \prime}
-2r\pu\pv\kk_{v}^{3 \prime}-(2M+18)r\kk_{v}^{3 \prime\prime}
-2\pu\pv \kk_{v}^{3}-(4M+12)\kk_{v}^{3 \prime}=0,\nn \\
&&-2r
\kk_{u}^{2 \prime\prime\prime}-\pu\pv\kk_{u}^{2 \prime}+\pu^{2}
\kk_{v}^{2 \prime}+(M-3)\kk_{u}^{2 \prime\prime}=0.
\eea

So far we have  presented the full linearized gauged fixed equations 
of motion of small fluctuations above an AdS vacuum of the MMG model. In what follows we shall
study their asymptotic solutions which can be used to determine  divergent terms 
appearing in the on shell action. This analysis,  in turns, may be used to fix 
the boundary terms needed to have a well-imposed variational principle.  

To proceed we solve the equations \eqref{TC} and \eqref{Lin.MMG.Final} order by order in 
$r$ near  the boundary at $r=0$.  Motivated by the pp-wave solution of the model 
 \cite{Alishahiha:2014dma}, we consider the following near boundary expansions for the 
 field $\kk$ 
\bea\label{kk}
&&\kk_{u}^{2} = \chi^{(0)}_{u}\log(r)+\varphi^{(0)}_{u}+\chi^{(1)}_{u}
r\log(r)+\varphi^{(1)}_{u}r+..., \nn \\
&&\kk_{v}^{3} =\frac{1}{2 r}\left(\chi^{(0)}_{v}\log(r)
+\varphi^{(0)}_{v}+\chi^{(1)}_{v}
r\log(r)+\varphi^{(1)}_{v}r+...\right) \nn \\
&&\kk_{v}^{2} = \tilde{\chi}^{(0)}_{v}\log(r)+
\tilde{\varphi}^{(0)}_{v}+\tilde{\chi}^{(1)}_{v}
r\log(r)+\tilde{\varphi}^{(1)}_{v}r+...
\eea
Plugging these expressions into the equations \eqref{TC} and \eqref{Lin.MMG.Final} and assuming 
 $(1+\sigma\alpha)\neq 0$ one may solve the equations as follows. Indeed from the equation \eqref{TC} at  $\log(r)/r$ order
one gets
\bea\label{As.1}
-\frac{M}{2\mu}\left(2\tilde{\chi}^{(1)}_{v}+\pv^{2}\chi^{(0)}_{u}
+\pu^{2}\chi^{(0)}_{v}-2\pu\pv \tilde{\chi}^{(0)}_{v}\right) =0,
\eea
while  at  $1/r^{2}$ order from  \eqref{TC} and the fifth equation
in \eqref{Lin.MMG.Final} one finds respectively
\bea\label{As.2}
-\frac{3M}{\mu}\tilde{\chi}^{(0)}_{v} =0,\;\;\;\;\;\;\;\;\;\;\;\;\;\;\;\;\;\;\frac{(M-1)}{\mu}\chi^{(0)}_{v} =0.
\eea
On the other hand at  $\log(r)$ order from the second and third equations 
of (\ref{Lin.MMG.Final}) one has 
\bea\label{As.3}
\frac{2(M-1)}{\mu}\left(\pv \chi^{(1)}_{u}-
\pu \tilde{\chi}^{(1)}_{v}\right) =0, \;\;\;\;\;\;\;\;\;\;\;
\frac{(M+1)}{\mu}\left(\pu \chi^{(1)}_{v} -
\pv \tilde{\chi}^{(1)}_{v}\right)=0.
\eea
 Moreover from the equation \eqref{TC} and the first, second, third, forth ,fifth and the 
 last equations of (\ref{Lin.MMG.Final})  at order 
$1/r$, one finds respectively
\bea\label{As.4}
&&\frac{M}{2\mu}\left(2\tilde{\chi}^{(1)}_{v}
-2\tilde{\varphi}^{(1)}_{v}-\pu^{2}\varphi^{(0)}_{v}
-\pv^{2}\varphi^{(0)}_{u}+2\pu\pv 
\tilde{\varphi}^{(0)}_{v}\right)=0,\;\;\;\;\;\;\;\;\;\;\;\;\;\;
\frac{4M}{\mu}\tilde{\chi}^{(0)}_{v} =0, \\
&&\frac{2}{\mu}\left([M+1]\pv \chi^{(0)}_{u}-[M-3]\pu
\tilde{\chi}^{(0)}_{v}\right)=0, \;\;\;\;\;\;\;\;\;\; \frac{1}{\mu}\left([M-1]\pu \chi^{(0)}_{v}-[M+3]\pv 
\tilde{\chi}^{(0)}_{v}\right)=0, \nn\\
&&-\frac{2M}{\mu}\tilde{\chi}^{(0)}_{v} =0,\;\;\;\;
\frac{1}{\mu}\left(-[M+1]\chi^{(1)}_{v}+\pv^{2}
\tilde{\chi}^{(0)}_{v}-\pu\pv \chi^{(0)}_{v}\right)=0,\;\;\;\;
\frac{2(M+1)}{\mu}\chi^{(0)}_{u} =0.\nn
\eea
Finally from the first, second, third, forth and the last equations of (\ref{Lin.MMG.Final}) at order 
of one, we find respectively 
\bea\label{As.5}
&&-\frac{2}{\mu}\left(2M \tilde{\chi}^{(1)}_{v}
+\pu^{2}\chi^{(0)}_{v}-\pv^{2}\chi^{(0)}_{u}\right)=0, \nn \\
&&\frac{1}{\mu}\left([2M-6]\pv \chi^{(1)}_{u}
-2[M+1]\pu \tilde{\chi}^{(1)}_{v}+2(M-1)[\pv
\varphi^{(1)}_{u}-\pu \tilde{\varphi}^{(1)}_{v}]\right)=0,\nn \\
&&\frac{1}{\mu}\left([1-M]\pv \tilde{\chi}^{(1)}_{v}
+(M+3)\pu \chi^{(1)}_{v}-(M+1)[\pv \tilde{\varphi}^{(1)}_{v}
-\pu \varphi^{(1)}_{v}]\right)=0,\nn \\
&&\frac{1}{\mu} \left(2M\tilde{\chi}^{(1)}_{v}
+\pu^{2}\chi^{(0)}_{v}-\pv^{2}\chi^{(0)}_{u}\right)=0,\nn \\
&&\frac{2}{\mu}\left([1-M]\chi^{(1)}_{u}-\pu^{2}
\tilde{\chi}^{(0)}_{v}+\pu\pv \chi^{(0)}_{u}\right) =0.
\eea
Now the aim is to solve these equations  to find different  components appearing in the 
asymptotic expansions \eqref{kk}. In particular for the cases of 
 $(\alpha=0,\mu \rightarrow \infty)$ and $(\alpha \neq 0, \mu\rightarrow \infty)$ where 
 $M \rightarrow \infty$ from  equations (\ref{As.1})-(\ref{As.5}) one arrives at 
\bea\label{As.Einstein}
&&\chi^{(0)}_{u}=\chi^{(0)}_{v}=
\tilde{\chi}^{(0)}_{v}=\chi^{(1)}_{u}=\chi^{(1)}_{v}=
\tilde{\chi}^{(1)}_{v}=0,\;\;\;\;\;\;
 \tilde{\varphi}^{(1)}_{v}=-\frac{1}{8}R[\varphi^{(0)}],\cr
&&\pv \tilde{\varphi}^{(1)}_{v}
-\pu \varphi^{(1)}_{v}=0,~~~~~\pv \varphi^{(1)}_{u}-
\pu \tilde{\varphi}^{(1)}_{v}=0,
\eea
where $R[\varphi^{(0)}] =4(\pu^{2}\varphi^{(0)}_{v}+\pv^{2}\varphi^{(0)}_{u}
-2\pu\pv \tilde{\varphi}^{(0)}_{v})$. It is worth noting that these results are in agreement 
with that obtained from metric formulation for 3D Einstein gravity \cite{deHaro:2000xn}. 
Note also that since the MMG model without 
the Chern-Simons term is the same as 3D Einstein gravity, these results show that adding  $\alpha$-term 
would not change the content of the 3D Einstein gravity.

On the other hand for  $\alpha =-\frac{2}{\mu}(1+\mu\sigma)$  where one has $M=-1$, from
the above equations,  one gets
\bea\label{As.Critical.MMG}
&&\chi^{(0)}_{v} =\tilde{\chi}^{(0)}_{v} =0,\;\;\;\;\;\;
\chi^{(1)}_{u}=-\frac{1}{2}\pu\pv \chi^{(0)}_{u}, \;\;\;\;\;\;
\tilde{\chi}^{(1)}_{v}=-\frac{1}{8}R[\chi^{(0)}],
\;\;\;\;\pu \chi^{(1)}_{v}=-\pv \tilde{\chi}^{(1)}_{v},\cr
&&\tilde{\varphi}^{(1)}_{v} =\tilde{\chi}^{(1)}_{v}
-\frac{1}{8}R[\varphi^{(0)}],\;\;\;\;\;\;\;\;
\pv 
\chi^{(1)}_{u}+\left(\pv \varphi^{(1)}_{u}
-\pu \tilde{\varphi}^{(1)}_{v}\right) =0.
\eea
Here we have used the  identity $R[\chi^{(0)}] = 4\pv^{2}\chi^{(0)}_{u}$.
%%%%%%%%%%%%%%%%%%%%%%%%%%%%%%%%%%%%%%%%%%%%
%%%%%%%%%%%%%%%%%%%%%%%%%%%%%%%%%%%%%%%%%%%%

\section{Variational Principle}\label{V.P}

In the previous section we studied asymptotic solutions of the linearized equations of motion 
of the MMG model. Of course the results rely on the fact that the model admits a well-imposed 
variational principle. This procedure 
requires proper boundary terms to make sure that all boundary terms can be consistently 
removed. In this section we would like to reexamine the variation of the action leading 
to the corresponding equations of motion.

To proceed let us consider the action of the MMG model 
\eqref{Modified.L.MMG} whose variation with respect to the fields 
$e, \Omega$ and $h$ are given by 
\bea\label{Var}
&& \delta_{e} L(e,\Omega, h) =E_{e}\cdot \delta e-D(\Omega)(h\cdot \delta e),\cr
&&\delta_{\Omega} L(e,\Omega, h) = E_{\Omega}\cdot \delta\Omega 
+D(\Omega)\left(\sigma e\cdot \delta\Omega-
\frac{1}{2\mu}(\Omega\cdot \delta\Omega-\alpha h\cdot \delta\Omega)\right),
\cr
&&\delta_{h} L(e,\Omega, h)= E_{h}\cdot \delta h+
D(\Omega)\left(-\sigma\alpha e\cdot \delta h -\frac{1}{2\mu}
(\alpha^2 h\cdot \delta h -\alpha \Omega\cdot \delta h)\right).
\eea 
Using the  Stokes' theorem\footnote{The normal vector to 
the boundary is $n_{\mu} = (-\frac{1}{2r},0,0)$.} the corresponding {
boundary terms} appearing in the above variation 
may be recast to the following form 
\bea\label{Var.FG}
&&\delta S|_{boundary} = \int_{\partial\mathcal{M}}d^{2}x
\epsilon^
{ij}\bigg(
\big[-C\mu\alpha\sigma e+\frac{C\alpha}{2}\Omega
-(1+\frac{C\alpha^{2}}{2})h\big]_{i a}
\delta e_{j}^{a}+\cr
&&~~+\big[\sigma e-\frac{1}{2\mu}\Omega
+\frac{\alpha}{2\mu}h\big]_{i a}\delta\Omega_{j}^{a}
+\big[-\alpha\sigma e+\frac{\alpha}{2\mu}\Omega
-\frac{\alpha^{2}}{2\mu}h\big]_{i a}\delta \pp_{j}^{a}\bigg),
\eea
where we have used $\delta h = C\mu\delta e+\delta \pp$ from Eq.(\ref{perturbation}). 

From the equations (\ref{v})  and (\ref{p})  one observes that the first and second
radial derivative of the driebein appear in the boundary terms. Hence, 
imposing the Dirichlet boundary condition is not enough to remove all
boundary terms. Indeed this is expected simply because even in the Einstein gravity 
 one needs to add a proper Gibbons-Hawking term.  In  our 
notation the corresponding  Gibbons-Hawking term  is
\be\label{G.H}
S_{GH} = 2\sigma \mathcal{K} = -\sigma\int_{\partial\mathcal{M}} d^{2}x
\epsilon^{ij}\tilde{e}_{i a}\tilde{\Omega}_{j}^{a},
\ee
where $\tilde{e}$ and $\tilde{\Omega}$ are boundary
driebein and  spin connection respectively. We note, however that,  as it is evident from Eq.{\eqref{Var.FG}}, the above Gibbons-Hawking term is not  sufficient to have a well-defined 
variational principle and, indeed,  more boundary terms are needed. Of course,  in general, 
determining these extra terms is not an easy task. Nonetheless 
for the model under consideration and for particular values of the parameter, one can
fix the boundary terms as follows. 

Before proceeding, it is illustrative to take a closer look at the boundary term containing the variation 
of $\delta\Omega_{j}^{a}$ in (3.2) which includes a variation of radial derivative of driebein. 
In the framework of the metric formulation it also contains a variation of radial derivative of 
boundary metric, $\delta\partial_{r}g_{ij}$. Let us first consider the case of  $\alpha=0$, 
where MMG reduces to TMG. In this case utilizing  the metric formulation of TMG it was shown that 
the contribution of Chern-Simons action (CS) to the coefficient of $\delta\partial_{r}g_{ij}$ vanishes 
for asymptotically locally AdS space-times (AlAdS) \cite{Kraus:2005zm}. However, one can see that 
in the driebein formulation the contribution of CS action to the coefficient of $\delta\Omega_{j}^{a}$ 
remains non-zero. This is, indeed, due to the fact that the variation of the CS action in metric and driebein formalisms are different by certain boundary terms (see Eq.(3.23) of 
Ref.\cite{Kraus:2005zm}).  Therefore to have a well-defined variational principle for TMG in driebein 
formalism, besides the standard Gibbons-Hawking term (\ref{G.H}),  one needs to add extra boundary terms.

It is worth noting that  recently the variational principle of TMG for AlAdS space-times in driebein 
formalism is studied in Ref.\cite{Grumiller:2015xaa}. There it is shown that just half of the 
standard Gibbons-Hawking term (\ref{G.H}) should be added to the action in order to have a well-defined 
variational principle. 
Actually as it is discussed in Refs.\cite{Miskovic:2006tm,Detournay:2014fva} there are two possible
ways to make the  variational principle  well defined in dreibein formalism for AlAdS space-times.
In the first way (the standard case)  one imposes the standard Dirichlet boundary condition on the
 driebein. In the second way we should impose mixed boundary conditions on both 
 boundary drebein and its first radial derivative. The study in Ref.\cite{Grumiller:2015xaa} is
 based on the  second option while in this paper we used the first option for
 AlAdS space-times. Of course 
 when $\alpha \neq 0$, the coefficients of $\delta\Omega_{j}^{a}$ and $\delta\pp_{j}^{a}$ remain 
 non-zero for AlAdS space-times.  In this case one can remove  the $\sigma$-term in coefficient of $
 \delta\Omega_{j}^{a}$ by adding the standard Gibbons-Hawking term (\ref{G.H}). Therefore when $
 \alpha \neq 0$,  one needs extra 
 boundary terms to have a well-defined variational principle for AlAdS space-times.
 
In this paper our main interest is the case $M = -1$, where the model exhibits logarithmic solutions. 
For this case to determine the solution we will have to know the values of driebein and its 
first radial derivative at the boundary. Therefore  a well-imposed variational principle 
occurs by imposing boundary conditions on driebein and its first radial derivative, simultaneously.  
Note that this is not the same as that in Ref.\cite{Grumiller:2015xaa}. This is  because in that work the variational principle is studied for AlAdS space-times while  the logarithmic solutions are not 
AlAdS space-times. 
With this treatment although $\delta\Omega_{j}^{a}$ contains the first radial derivative, the corresponding term does not invalidate the modified variational principle. Nonetheless one still needs to deal with the last term in Eq.(3.2) that contains the variation $\delta \pp_{i}^{a}$ . 

Actually we note that different components of  $\pp_{i}^{a}$  contain  the second 
radial derivative of driebein and thus have a potential to invalidate even the 
modified variational principle. More precisely one has 
\be\label{Danger.1}
\delta\pp_{u}^{2} =\frac{4r^{2}}{\mu(1+\sigma\alpha)^{2}}\delta\kk_{u}^{2 \prime\prime},
~~~~~~\delta\pp_{v}^{3} =\frac{4r^{2}}{\mu(1+\sigma\alpha)^{2}}\delta\kk_{v}^{3 \prime\prime}
+\frac{8r}{\mu(1+\sigma\alpha)^{2}}\delta\kk_{v}^{3 \prime}.
\ee 
Therefore the corresponding terms which could invalidate the modified variational principle
are
\be\label{Danger.2}
\bigg[-\alpha\sigma e+\frac{\alpha}{2\mu}\Omega
-\frac{\alpha^{2}}{2\mu}h\bigg]_{v 2}\delta\pp^2_u-
\bigg[-\alpha\sigma e+\frac{\alpha}{2\mu}\Omega
-\frac{\alpha^{2}}{2\mu}h\bigg]_{u 3}\delta\pp^3_v.
\ee
By making use of  the asymptotic analysis  given in the equation   
(\ref{As.Critical.MMG}) we get the following asymptotic behavior for $\kk$
\bea
&&\kk_{u}^{2} = \chi^{(0)}_{u}\log(r)+\varphi^{(0)}_{u}+\chi^{(1)}_{u}
r\log(r)+\varphi^{(1)}_{u}r+\cdots\,, \cr
&&\kk_{v}^{3} =\frac{1}{2 r}(\varphi^{(0)}_{v}+\chi^{(1)}_{v}
r\log(r)+\varphi^{(1)}_{v}r+\cdots)\, , \cr
&&\kk_{v}^{2} = \tilde{\varphi}^{(0)}_{v}+\tilde{\chi}^{(1)}_{v}
r\log(r)+\tilde{\varphi}^{(1)}_{v}r+\cdots\, .
\eea
Plugging these asymptotic expressions in (\ref{Danger.2}) and using Eqs.(\ref{Danger.1}), (\ref{v}) and (\ref{p}) one arrives at 
\be
\mathcal{O}(r^2)\delta\kk_{u}^{2 \prime\prime} 
+\mathcal{O}(r)\delta\kk_{v}^{3 \prime\prime},
\ee 
showing that the boundary terms  (\ref{Danger.2}) vanish near the  boundary and 
thus would not  invalidate the
modified variational principle. To summarize, we note that  for the  MMG model at the critical 
point one just needs to  add the usual
Gibbons-Hawking term to have a well-defined modified variational principle. Of course to
find a finite on-shell action we will have to
add certain counterterms which we will discuss in the following section.

%%%%%%%%%%%%%%%%%%%%%%%%%%%%%%%%%%%%%%%%%%%%%%%%%%%%%%%
\section{MMG On-Shell Action}

In this section we would like to compute the on-shell action from which one may evaluate  
 the expectation values of boundary operators using AdS/CFT correspondence. 
 In this context, higher order 
correlation functions can also be obtained by differentiating the on-shell action 
with respect to the boundary sources. 

Using the results of the previous section the total action of MMG is given by 
\be
S = S_{MMG} + S_{GH},
\ee
which is guarantied that the model has a well defined variational principle.  Nonetheless
it is however important to note that even with the Gibbons-Hawking term the 
on-shell action might  diverge and in order to  remove the divergent terms one
needs to add proper boundary counterterms. This is the aim of this section to 
determine the  divergent terms and then  the proper boundary counterterms.
% which remove the divergences.

%\subsection{Regularized MMG On-Shell Action}

To regularize  the on-shell action one can introduces a UV cut-off at $r \geq \epsilon$ and finds
the corresponding  boundary terms at $r=\epsilon$. Let us  expand the   
regularized on-shell action in power of the fields as follows
\be
S_{on-shell}\big{|}_{reg} = S^{(0)}+S^{(1)}+S^{(2)}+{\cal O}({\rm {fields}}^3),
\ee
where by $S{^{(0)}}, S^{(1)}$ and $S^{(2)}$ are zeroth order (background), linearized and 
quadratic actions of the perturbations, respectively. It is then straightforward, though tedious, to compute divergent terms of each part. The final results may be summarized as follows 
\bea\label{Div.OnShell}
&&S_{div}^{(0)} = \int
d^{2}x\;\frac{1}{\epsilon}\mathcal{S}^{(0)}_{1/\epsilon},\\
&&S_{div}^{(1)}= \int
d^{2}x\left(\log^{2}(\epsilon)\mathcal{S}^{(1)}_{\log^{2}(\epsilon)} 
+\log(\epsilon)\mathcal{S}^{(1)}_{\log(\epsilon)} 
+\frac{1}{\epsilon}\mathcal{S}^{(1)}_{1/\epsilon}\right),\cr
&&S_{div}^{(2)}=\int
d^{2}x\left(\log^{3}(\epsilon)\mathcal{S}^{(2)}_{\log^{3}(\epsilon)}
+\log^{2}(\epsilon)\mathcal{S}^{(2)}_{\log^{2}(\epsilon)}
+\frac{\log(\epsilon)}{\epsilon}\mathcal{S}^{(2)}_{\log(\epsilon)/\epsilon}
+\log(\epsilon)\mathcal{S}^{(2)}_{\log(\epsilon)}
+\frac{1}{\epsilon}\mathcal{S}^{(2)}_{1/\epsilon}\right).\nn
\eea
The explicit expressions of the coefficients in the above equation are presented in the
 appendix \ref{Div}.  In what follows we will study these divergent terms for the cases
  $(\alpha=0,\mu \rightarrow \infty),
(\alpha \neq 0, \mu\rightarrow \infty)$ and
$(\alpha =-\frac{2}{\mu}(1+\mu\sigma))$, separately. 

Actually for the cases $(\alpha=0,\mu \rightarrow \infty)$ 
and $(\alpha \neq 0, \mu\rightarrow \infty)$, from the explicit expressions given in  the appendix \ref{Div} one gets\footnote{
To be precise,  for MMG in these two cases we have used Eqs.(\ref{As.Einstein}) 
instead of (\ref{As.Critical.MMG}). Actually in these cases 
the equations in  (\ref{As.Critical.MMG})
reduce to that in (\ref{As.Einstein}) except for the last equations. },
%We note, however, 
%the corresponding equations do not cause any problem simply because in order 
%to find the  divergent terms we do not need  these  equations.},
\be\label{Div.Ein}
\mathcal{S}^{(0)}_{1/r} = \sigma, ~~~\mathcal{S}^{(1)}_{\log(r)} = -
\frac{\sigma}{4}R[\varphi^{(0)}],
~~~\mathcal{S}^{(1)}_{1/r} = 2\sigma \tilde{\varphi}^{(0)}_{v},~~~
\mathcal{S}^{(2)}_{1/r} = -\sigma(\varphi^{(0)}_{u}\varphi^{(0)}_{v}-
\tilde{\varphi}^{(0) 2}_{v}).
\ee
Hence, the proper boundary counterterms needed to remove these divergent terms are
\be
S_{ct} = 2\sigma \int
d^{2}x~\tilde{e}+\frac{\sigma}{2}\int d^{2}x
R[\tilde{e}]\log(\epsilon),
\ee 
where $R[\tilde{e}] = -\frac{1}{2}\epsilon^{ij}\partial_{i}(
\frac{1}{\tilde{e}}\epsilon^{kl}\tilde{e}_{j}^{a}\partial_{k}
\tilde{e}_{la})$ is scalar curvature of 2D boundary geometry.  Then the renormalized
on shell action becomes
\be\label{Ren.Ein}
S_{ren} = \sigma\int d^{2}x
\left(\varphi^{(0)}_{u}\varphi^{(1)}_{v}+\varphi^{(0)}_{v}\varphi^{(1)}_{u}
+\frac{1}{4}\tilde{\varphi}^{(0)}_{v}R[\varphi^{(0)}]\right),
\ee
in agreement with Ref.\cite{deHaro:2000xn} where the corresponding renormalized on-shell
action has been found in the metric formalism. It is worth noting that the form of divergences
as well as the renormalized on-shell action for both cases
$(\alpha=0,\mu \rightarrow \infty)$ and 
$(\alpha \neq 0, \mu\rightarrow \infty)$  are the same; indicating that these two models might
have the same physical contents.

On the other hand using the expressions presented in the appendix \ref{Div}
for $\alpha =-\frac{2}{\mu}(1+\mu\sigma)$ one arrives at\footnote{
For this case we have $\Lambda_{0} =\frac{1}{2}\mu(\mu\sigma+3)$
and $C=\frac{1}{2}$.}
\bea\label{Div.MMG}
&&\mathcal{S}^{(0)}_{1/r} = \frac{1}{4\mu}(5\mu\sigma +1),~~~~
\mathcal{S}^{(1)}_{\log^{2}(r)} = -
\frac{3}{32}\frac{\mu\sigma -1}{\mu}R[\chi^{(0)}], \\
&&\mathcal{S}^{(1)}_{\log(r)} =-\frac{3\sigma\mu^{3}+11\mu^{2}-12}
{16\mu(2+\mu\sigma)^{2}}R[\chi^{(0)}]
-\frac{3}{16}\frac{\mu\sigma -1}{\mu} R[\varphi^{(0)}],~~~~
\mathcal{S}^{(1)}_{1/r} = \frac{5\mu\sigma+1}{2\mu}\tilde{\varphi}^{(0)}_{v}, \cr
&&\mathcal{S}^{(2)}_{\log^{3}(r)} = \frac{\mu\sigma +1}{12\mu} 
\chi^{(0)}_{u}\chi^{(1)}_{v},\;\;\;\;\;\;\;\;\mathcal{S}^{(2)}_{\log(r)/r} =-\frac{5\mu\sigma +1}
{4\mu}\chi^{(0)}_{u}\varphi^{(0)}_{v},
\cr
&&
\mathcal{S}^{(2)}_{\log^{2}(r)} =\frac{\mu\sigma +1}{8\mu}
\left(\varphi^{(0)}_{u}\chi^{(1)}_{v}+\chi^{(0)}_{u}\varphi^{(1)}_{v}
-\chi^{(0)}_{u}\pv^{2}\tilde{\varphi}^{(0)}_{v}+\frac{3}{2}\chi^{(0)}_{u}
\pu\pv \varphi^{(0)}_{v}\right), \cr
&&
\mathcal{S}^{(2)}_{\log(r)} =\frac{(\mu\sigma+1)}
{\mu(2+\mu\sigma)^{2}}\bigg[ \frac{3\sigma\mu^{3}+25\mu^{2}
+16\mu\sigma -4}{1+\mu\sigma}\chi^{(0)}_{u}\chi^{(1)}_{v}
-2(\chi^{(0)}_{u}\varphi^{(1)}_{v}-\varphi^{(0)}_{u}\chi^{(1)}_{v})\cr
&&\;\;\;\;\;\;\;\;\;\;\;\;\;+\frac{1}{4}(2+\mu\sigma)^{2}
(\varphi^{(0)}_{u}\varphi^{(1)}_{v}-3\varphi^{(0)}_{v}\varphi^{(1)}_{u}
-\frac{1}{4}\tilde{\varphi}^{(0)}_{v}R[\varphi^{(0)}])\nn\\
&&\;\;\;\;\;\;\;\;\;\;\;\;\;-\frac{3}{4}(\mu^{2}+4\mu\sigma -4)\chi^{(0)}_{u}\pv^{2}\tilde{\varphi}^{(0)}_{v}
+\frac{1}{2}(\mu^{2}
+4\mu\sigma +2) \chi^{(0)}_{u}\pu\pv \varphi^{(0)}_{v}\bigg],\cr
&&\mathcal{S}^{(2)}_{1/r} = \frac{7\sigma\mu^{3}+55\mu^{2}
+48\mu\sigma+4}{4\mu (2+\mu\sigma)^{2}}\chi^{(0)}_{u}\varphi^{(0)}_{v}
-\frac{5\mu\sigma +1}{4\mu}(\varphi^{(0)}_{u}\varphi^{(0)}_{v}-\tilde{\varphi}^{(0) 2}_{v}
).\nn
\eea
Before proceeding to write  proper counterterms to remove the above divergent terms, let us 
explore certain features of these divergent terms.  We note that although for  
$\sigma=-1, \mu=1$ the  equations of motion of
critical MMG reduce to that of TMG at the  critical point, the above divergent terms which are
obtained in the driebein formalism do not reduce to divergent terms obtained from the  metric formalism\cite{Skenderis:2009nt}. This may be understood as follows.
Actually the Chern-Simons action in the  driebein formalism
 differs from Chern-Simons action in the metric formalism by certain  terms whose  
variations are  boundary terms. Therefore their on-shell values should also be 
different by these  boundary terms (see {\it e.g}  \cite{Kraus:2005zm}).

We note also that the above  divergent terms  contain several terms such as 
$\chi^{(0)}_{u}\chi^{(1)}_{v}$, whose corresponding counterterms needed to remove them 
include the driebein as well as its radial derivative.
Of course these  types of boundary counteretrms are not  allowed 
for pure 3D Einstein gravity,  though they are consistent with  MMG at the critical point, as  
discussed in section \ref{V.P}.

Finally as it is evident from  expressions of (\ref{Div.MMG}) for a fixed value of
$\mu$, the divergent terms change as one changes  $\sigma=1$  to $\sigma=-1$. Therefore the
 corresponding counterterms should be considered separately. 
 Actually it can be seen that in order to remove divergent terms coming from the zeroth and 
first order of the perturbations  one needs to add the following counterterms 
\be\label{sigma.m1.1}
S^{1}_{ct} =\int
d^{2}x\left(- \frac{5\sigma \mu +1}{2\mu}~\tilde{e}-\frac{3}{8}\frac{\mu +1}{\mu}
R[\tilde{e}]\log(\epsilon)\right),
\ee 
while the proper counterterms required to remove the divergent terms at second order 
of perturbations for $\sigma=-1$
are given by 
\bea\label{sigma.m1.2}
&&S^{2}_{ct} = \int
d^{2}x\bigg[a_{1}\pp\star\kk
+a_{2}\kk\centerdot\kk +a_{3}\kk\centerdot\vv +
a_{4}\vv \centerdot\vv +a_{5}\pp\star\vv +a_{6}\pp\centerdot\vv
-4a_{7}\kk^{ij}\bar{\Box}\kk_{ij} \cr
&&+\left(
b_{1}\pp\star\kk +b_{2}\pp\centerdot\kk +b_{3}\kk\star\vv +b_{4}\kk\centerdot\vv +
b_{5}\pp\star\vv +b_{6}\pp\centerdot\vv +
b_{7}\pp\centerdot\pp +b_{8}\tilde{e}^{(2)}\right)\log(\epsilon)\cr
&&+\left(c_{1}\pp\star\kk +c_{2}\pp\centerdot\kk
+c_{3}\pp\centerdot\vv +c_{4}\pp\centerdot\pp +\frac{1}{2}c_{5}
\mathcal{K}_{ij}\mathcal{K}^{ij}\right)\log^{2}(\epsilon)
+d_{1}\pp\centerdot\pp \log^{3}(\epsilon)\bigg].\nonumber\\
\eea
However, for $\sigma=1$ they are 
\bea\label{sigma.1.2}
&&S^{2}_{ct} = \int d^{2}x\bigg[a_{1}\pp\star\kk
+a_{2}\kk\centerdot\kk +a_{3}\pp\centerdot\kk +
a_{4}\kk \centerdot\vv +a_{5}\vv\centerdot\vv +a_{6}\pp\star\vv
-4a_{7}\kk^{ij}\bar{\Box}\kk_{ij} \cr
&&
+(b_{1}\pp\star\kk +b_{2}\pp\centerdot\kk +b_{3}\kk\star\vv +b_{4}\kk\centerdot\vv +
b_{5}\pp\star\vv +b_{6}\pp\centerdot\vv 
+b_{7}\tilde{e}^{(2)}+\frac{1}{2}b_{8}\mathcal{K}_{ij}
\mathcal{K}^{ij})\log(\epsilon)\cr
&&+\left(c_{1}\pp\star\kk +c_{2}\pp\centerdot\kk
+c_{3}\pp\centerdot\vv +c_{4}\pp\centerdot\pp +\frac{1}{2}c_{5}
\mathcal{K}_{ij}\mathcal{K}^{ij}\right)\log^{2}(\epsilon)
+d_{1}\pp\centerdot\pp \log^{3}(\epsilon)
\bigg].\nonumber\\
\eea
Here $``\star''$ and $``\centerdot''$ products are defined by  
$\mathbf{A}\star\mathbf{B}=\epsilon^{ij}\mathbf{A}_{i}^{a}\mathbf{B}_{j a},\; 
\mathbf{A}\centerdot\mathbf{B}=\eta^{ij}\mathbf{A}_{i}^{a}
\mathbf{B}_{j a}$, respectively.   Moreover $\kk_{ij}
=\bar{e}_{i}^{a}\kk_{j a}$, $\bar{\Box}$ is defined by 2D flat metric,  $\tilde{e}^{(2)}$ is the second 
order in perturbations of boundary driebein determinant and 
$\mathcal{K}_{ij}$ is the first order in perturbations of
boundary extrinsic curvature.
%\footnote{
%From explicit form of $\vv_{i}^{a},\pp_{i}^{a}$, which can be calculated from (\ref{v})
%and (\ref{p}), one can see that they are determined completely 
%by perturbations of boundary driebein.}. 
The explicit values of the coefficients in the above expressions
for $\sigma=-1$ and $\sigma=1$ are given in the  appendix \ref{EVC}.

Adding these  counterterms to the on-shell action, all divergent terms given in 
the equation (\ref{Div.MMG}) will be removed leading  to the following  renormalized on-shell
action\footnote{The renormalized on-shell action also contains local terms whose 
contributions to the correlation functions are contact terms. Note that  
these local terms can be changed
by adding finite local functions. The explicit forms of these local terms are presented in  
 the appendix \ref{LFO}. } 
\be\label{S.Ren.m1}
S_{ren}\!\!=\!\! \int\! d^{2}x \bigg[\!\frac{5\mu +\sigma+8}{2\mu}
\chi^{(0)}_{u}\chi^{(1)}_{v}\!-\!\frac{4\mu}{(\sigma\mu +2)^{2}}
\chi^{(0)}_{u}\varphi^{(1)}_{v}\!-\!\frac{4\mu}{(\sigma\mu +2)^{2}}
\varphi^{(0)}_{u}\chi^{(1)}_{v}\!-\!\frac{4}{\mu}\varphi^{(0)}_{v}\varphi^{(1)}_{u}\!
\bigg].
\ee
It is worth noting that the above counterterms contain the first and second 
radial derivatives of the perturbations which in general could destroy the 
variational principle. We note, however, that following our discussions in the previous 
section the first radial derivative can be accommodated by the modified boundary 
condition at the critical point and moreover it can be shown that those terms
with the second radial derivative vanish as one approaches the boundary. 
Therefore the resultant counterterms are consistent with our variational principle.

%%%%%%%%%%%%%%%%%%%%%%%%%%%%%%%%%%%%%%%%%%%%%%%%%%%%%
\section{Correlation Functions}

Having found the renormalized on shell action it is now possible to compute correlation 
functions of the holographic stress tensor. Following AdS/CFT correspondence the non-normalizable 
modes of a bulk field can be identified as a source for a dual operator. Moreover  the $n^{th}$ correlation functions of the dual operator  can be evaluated by taking $n^{th}$ functional derivative 
of the on shell action with respect to the source. 

In the present case the corresponding sources  are  $\varphi^{(0)}_{u},\varphi^{(0)}_{v}
,\tilde{\varphi}^{(0)}_{v}$ and $\chi^{(0)}_{u}$ and therefore  one point functions of
the associated dual operators are given by\footnote{The overall factor in action is $1/16\pi G$ and for  normalization of one-point and
two-point functions we have used the same 
convention as in Ref.\cite{Skenderis:2009nt}.  Note also that the sign difference between the 
definition  $\langle T_{ij}\rangle$ and  $\langle t_{ij}\rangle$
may  be understood from the changing the sign of perturbations
by raising or lowering the indices.}.
\be\label{One.Ponit.Def}
\langle T_{ij}\rangle = 2\pi \left(\bar{e}_{i a}
\frac{\delta S_{ren}}{\delta e^{(0) j}_{a}} + i \leftrightarrow j\right),~~~
~~\langle t_{ij}\rangle = -2\pi \left(\bar{e}_{i a}
\frac{\delta S_{ren}}{\delta \chi^{(0) j}_{a}} + i \leftrightarrow j\right).
\ee
Here we have use the fact that  $e^{(0) v}_{3}\equiv 2\varphi^{(0)}_{u},
e^{(0) u}_{2}\equiv 2\varphi^{(0)}_{v}
,e^{(0) u}_{3}\equiv 2\tilde{\varphi}^{(0)}_{v}$ and 
$\chi^{(0) v}_{3} \equiv 2\chi^{(0)}_{u}$. 

Using the explicit form of the on-shell action given in the equation 
\eqref{S.Ren.m1} one can evaluate the one point functions as follows
\bea\label{One.Point}
&&\langle T_{uu}\rangle =-\frac{1}{2\mu G}\varphi^{(1)}_{u}+{\rm local},~~~~
\langle T_{vv}\rangle = -\frac{1}{2G}\frac{\mu}{(\mu\sigma +2)^{2}}
\chi^{(1)}_{v}+{\rm local},~~~~\langle T_{uv}\rangle ={\rm local},\cr
&&~~~~~~~~~~~~~~~~\langle t_{vv}\rangle = \frac{1}{8G}\bigg(
-\frac{5\mu +\sigma +8}{2\mu}\chi^{(1)}_{v}+\frac{4\mu}{(\mu\sigma +2)^{2}}
\varphi^{(1)}_{v}\bigg)+{\rm local},
\eea
where the  "local"  terms   are local functions of sources whose contributions
to  two-point functions are  contact terms. The explicit expressions
of these local terms are given  in the appendix \ref{LTO}.

Using AdS/CFT procedure the higher point correlation functions could also be computed as
follows
\bea\label{Two.Point.def}
\langle T_{ij} \cdots\rangle &=& -2\pi i \left(\bar{e}_{i a}
\frac{\delta }{\delta e^{(0) j}_{a}} + i \leftrightarrow j\right)
\langle \cdots \rangle,\cr
\langle t_{ij} \cdots\rangle &=& 2\pi i \left(\bar{e}_{i a}
\frac{\delta }{\delta \chi^{(0) j}_{a}} + i \leftrightarrow j\right)
\langle \cdots \rangle .
\eea

It is important to note that in order to find the higher order correlation functions
one should know the explicit dependence of the response functions on the sources.
Actually this  explicit form  cannot be found from the asymptotic analysis of the equations of motion.
Indeed in order to find this dependence one needs
 to solve exactly the linearized equations of motion (\ref{Lin.MMG.Final}) and (\ref{TC}) in the bulk. 
 In general it is not an easy task to solve the corresponding  
 linearized equations and it might not even have an
 explicit analytic solution. Fortunately in the preset case these equations at the 
 critical point $M=-1$ can be solve analytically.
 %%%%%%%%%%%%%%%%%%%%%%%%%%%%%%%%%%%%%%%%%%%%%%%%%%%%
 
%\subsection{Exact Critical Solutions}
Actually at critical value $M=-1$, the equations (\ref{TC}) and  (\ref{Lin.MMG.Final}) reduce to
\bea\label{critical.eq}
&&\frac{1}{2}\pv^{2}\kk_{u}^{2}-\pu\pv \kk_{v}^{2}
+\kk_{v}^{2 \prime} +r\pu^{2}\kk_{v}^{3}-2r
\kk_{v}^{2 \prime\prime} =0\cr
&&-\frac{1}{2}\pv^{2}\kk_{u}^{2 \prime}+\pu^{2}\kk_{v}^{3}
-\kk_{v}^{2 \prime\prime}
+r\pu^{2}\kk_{v}^{3 \prime} =0,\cr
&&2r\pv \kk_{u}^{2 \prime\prime}-\pu\kk_{v}^{2 \prime}
+2r\pu\kk_{v}^{2 \prime\prime}+\pv\kk_{u}^{2 \prime}
+\pv\kk_{u}^{2 \prime}-\pu \kk_{v}^{2 \prime}=0 \cr
&&-4r^{2}\pu\kk_{v}^{3 \prime\prime} +\pv\kk_{v}^{2 \prime}-2r\pv\kk_{v}^{2 \prime\prime}
+2r\pu\kk_{v}^{3 \prime}-10r\pu\kk_{v}^{3 \prime}-\pv\kk_{v}^{2 \prime}
=0 \cr
&&2r^{2}\pu^{2}\kk_{v}^{3 \prime}-r\pv^{2}\kk_{u}^{2 \prime}+2r\pu^{2}
\kk_{v}^{3}-2r\kk_{v}^{2 \prime\prime} =0, \cr
&&-4r^{2}\kk_{v}^{3 \prime\prime\prime} +\pv^{2}\kk_{v}^{2 \prime}
-2r\pu\pv\kk_{v}^{3 \prime}-16 r\kk_{v}^{3 \prime\prime}
-2\pu\pv \kk_{v}^{3}-8\kk_{v}^{3 \prime} =0,\cr
&&-2r\kk_{u}^{2 \prime\prime\prime}-\pu\pv\kk_{u}^{2 \prime}+\pu^{2}
\kk_{v}^{2 \prime}-4\kk_{u}^{2 \prime\prime} =0.
\eea
One observes that with the identification  
$\kk_{u}^{2}=h_{uu},~\kk_{v}^{2}=h_{uv},~\kk_{v}^{3}=\frac{1}{2r}h_{vv}$
the above equations are the same as that given in the equation (6.19) of  \cite{Skenderis:2009nt} 
at critical value $\mu =1$. Similar observation has also been made  
in Ref.\cite{Alishahiha:2014dma} where the linearized equations of motion of 
MMG have been studied in the transverse traceless gauge.
Note that Eq.(\ref{critical.eq}) contains seven equations. However, by the above identification the second and fifth equations
in Eq.(\ref{critical.eq}) become identical.
Note that this identification can also be understood as follows.
In fact at the linearized  level from the  definition $g_{\mu\nu}=\eta_{a b}
e_{\mu}^{a}e_{\nu}^{b}$ and  in FG gauge one finds
\be
\frac{1}{r}h_{ij} = \kk_{i}^{a}\bar{e}_{j a} +\bar{e}_{i}^{a}\kk_{j a},
\ee
which reduces to the above identification using Eqs.(\ref{drie.FG.AdS}) and (\ref{k}).

By making use of this observation one can utilize the full solution of the TMG model
obtained in  \cite{Skenderis:2009nt} whose near boundary expansions are
\bea\label{Full.sol}
&&\kk_{u}^{2} = \chi^{(0)}_{u} \log(r)+\varphi^{(0)}_{u}-\frac{1}{2}
r\log(r)\pu\pv \chi^{(0)}_{u} +r\left((\pv)^{-1}\pu \tilde{\varphi}^{(1)}_{v}+
\frac{4\gamma -3}{2}\pu\pv \chi^{(0)}_{u}\right)+\cdots ,\cr
&&\kk_{v}^{3} = \frac{1}{2r} ( \varphi^{(0)}_{v} +\frac{1}{2}r\log(r)(\pu)^{-1}\pv^{3}
\chi^{(0)}_{u}\cr
&&\;\;\;\;\;\;\;\;+r\bigg[(\pu)^{-1}\pv \tilde{\varphi}^{(1)}_{v}+\bigg(2\gamma -1+2\log(\frac{q}{2})\bigg)
(\pu)^{-1}\pv^{3} \chi^{(0)}_{u}\bigg]+\cdots,\cr
&&\kk_{v}^{2} = \tilde{\varphi}^{(0)}_{v}-\frac{1}{2}r\log(r)\pv^{2}
\chi^{(0)}_{u}+r\tilde{\varphi}^{(1)}_{v}+\cdots,
\eea
where
\be
\tilde{\varphi}^{(1)}_{v} = -\frac{1}{2}\pv^{2}\chi^{(0)}_{u}-
\frac{1}{2}\pv^{2}\varphi^{(0)}_{u}-\frac{1}{2}\pu^{2}
\varphi^{(0)}_{v}+\pu\pv \tilde{\varphi}^{(0)}_{v}.
\ee
Here $\chi^{(0)}_{u},\varphi^{(0)}_{u},\varphi^{(0)}_{v},
\tilde{\varphi}^{(0)}_{v}$ are
arbitrary functions of the boundary coordinates $(u,v)$,  $q\equiv\sqrt{-4\pu\pv}$ and
$(\partial_{i})^{-1}$ denotes an  integration with respect to $x^{i}$. Note that 
as far as the two-point function of stress tensor are concerned, the above 
explicit expansions of the full solution are enough. 

We have now  all ingredients to compute  two point functions of the stress tensor  $T$ and its 
logarithmic partner $t$. To proceed one needs to express the one point functions (\ref{One.Point})
as functions of sources. This can be achieved by substituting
the full solution (\ref{Full.sol}) into the  equation (\ref{One.Point})
leading to 
\bea\label{One.Point.Simp}
&&\langle T_{uu}\rangle =\frac{1}{4\mu G}(\pv)^{-1}\pu^{3}
\varphi^{(0)}_{v}+{\rm local},\;\;\;\;\;\;\;
\langle T_{vv}\rangle = -\frac{1}{4G}\frac{\mu}{(\mu\sigma +2)^{2}}
(\pu)^{-1}\pv^{3}\chi^{(0)}_{u}+{\rm local},\cr
&&\langle T_{uv}\rangle ={\rm local},\cr
&&\langle t_{vv}\rangle =\frac{1}{2G}\frac{\mu}{(\mu\sigma +2)^{2}}
\bigg(\big[-\frac{(5\mu +\sigma +8)(\mu\sigma +2)^{2}}{16\mu^{2}}
-\frac{3}{2}+2\gamma +2\log(\frac{q}{2})\big]
(\pu)^{-1}\pv^{3}\chi^{(0)}_{u}\cr
&&~~~~~~-\frac{1}{2}(\pu)^{-1}\pv^{3}\varphi^{(0)}_{u}\bigg) +{\rm local}.
\eea
From  the identities 
%By doing the variation of one-point functions 
%(\ref{One.Point.Simp}) according to (\ref{Two.Point.def})
%and using identities
\bea
&&\frac{1}{\pu\pv}\delta^{2}(u,v) =\frac{i}{2\pi}\log(m^{2}uv),\;\;\;\;\;\;\pu^{4}[\frac{i}{2\pi}\log(m^{2}uv)] =-\frac{3i}{\pi}\frac{1}{u^{4}},\cr
&&\log(q)(\pu)^{-1}\pv^{3}\delta^{2}(u,v) =
\frac{i}{4\pi}[-\frac{11}{v^{4}}+\frac{6}{v^{4}}\log(m^{2}uv)],
\eea
one can compute derivatives of one point functions with respect to the sources to find 
two point functions as follows
\bea\label{Two.Point}
&&\langle T_{uu}T_{uu}\rangle = -\frac{3}{2\mu G}\frac{1}{u^{4}},~~~~~~~~~
\langle T_{vv}t_{vv} \rangle =-\frac{3}{2G}\frac{\mu}{(\mu\sigma +2)^{2}}
\frac{1}{v^{4}},\cr
&&\langle t_{vv}t_{vv}\rangle =\frac{3}{G}\frac{\mu}{(\mu\sigma +2)^{2}}
\frac{\log(m^{2}uv)}{v^{4}}-\frac{1}{G}\frac{B}{v^{4}},
\eea
and the correlation of other components are zero. Here $B$ is a constant which can be 
removed by a field redefinition, $t_{vv} \rightarrow t_{vv} -
B\frac{(\mu\sigma +2)^{2}}{3\mu}T_{vv}$.

It is shown that the asymptotic symmetry algebra of MMG model
consists of two copies of the Virasoro algebra with  central charges
\cite{Bergshoeff:2014pca}
\be
c_{\pm} = \frac{3}{2G}(\sigma \pm \frac{1}{\mu} + \alpha C),
\ee
which at critical point $\alpha = -\frac{2}{\mu}(1+\mu\sigma)$
they become
\be\label{c.MMG.critical}
c_{+} =0,~~~~~c_{-} =-\frac{3}{\mu G},
\ee
which is consistent with the results given in Eq.(\ref{Two.Point}). Let us recall that in our notation
the correlation functions of  a LCFT are  \cite{Gurarie:2004ce}
% value of $c_{-}$ in is agreement with two-point function of $T_{uu}$
%in (\ref{Two.Point}). Moreover, by comparing  with correlation functions 
%in a LCFT which is given by
 \bea
&&\langle T_{uu}T_{uu}\rangle = \frac{c_{-}}{2 u^{4}},~~~~~~~~~\langle T_{vv}T_{vv}\rangle = \frac{c_{+}}{2 v^{4}},~~~~~~~~~
\langle T_{vv}t_{vv} \rangle =\frac{b_{+}}{2 v^{4}},\cr
&&\langle t_{vv}t_{vv}\rangle =-\frac{\log(m^{2}uv)[c_{+}\log(m^{2}uv)+2b_{+}]}{2v^{4}}.
\eea
Comparing these expressions with that of Eq.\eqref{Two.Point} one finds 
$b_{+}=-\frac{3}{G}\frac{\mu}{(\mu\sigma+2)^{2}}$ in agreement with  \cite{Alishahiha:2014dma}.

\section{Conclusions}

In this paper we have studied  holographic renormalization
of  recently proposed 3D gravitational theory
known as  MMG \cite{Bergshoeff:2014pca} in the driebein formalism. In particular 
we have considered the model at the critical point $M=-1$ where it exhibits 
a new logarithmic solution.

To study holographic renormalization of the model for the logarithmic solution one needs
to fix both the  driebein and
its first radial derivative at the boundary. This, in turns, indicates that the driebein and 
its derivative correspond to sources of two operators with dimension two in the holographic dual
field theory. Indeed the dual field theory would be a logarithmic CFT  and the corresponding 
dual operators are holographic stress tensor and its logarithmic partner. Using the 
holographic renormalization method we have evaluated the correlation functions among 
different components of these operators which confirm that the dual theory is indeed a
logarithmic CFT. 

It is important to note that in order to compute the correlation functions of holographic 
stress tensor and its logarithmic partner one has to make sure that the model has
a well-defined variational principle and moreover the  on-shell  action for the corresponding 
solution is finite.  In general to meet these two conditions one has to add proper boundary 
terms which  include both the Gibbons-Hawking terms as well as certain counterterms. 
We have seen that as far as the model is concerned at critical point, the standard Gibbons-Hawking term is enough to have a well-defined
variational principle. The full expressions
for counterterms have also been presented.

It is also worth noting that using the holographic renormalization we have  seen 
that the model at different limits  ($\alpha = 0, \mu \rightarrow \infty $) and
 ($\alpha \neq 0,\mu \rightarrow \infty$) has the same structure showing that
 adding the $\alpha$ term to the Einstein-Hilbert would not change the content of the theory.
%%%%%%%%%%%%%%%%%%%%%%%%%%%%%%%%%%%%%%

%%%%%%%%%%%%%%%%%%%%%%%%%%%%%%%%
\section*{Acknowledgments}

We would like to thank Wout Merbis for useful discussions.

\section*{Appendix}

\appendix

\section{Explicit form of the action}
\vspace{-1cm}
\bea
&& S_{MMG} = \int_{\mathcal{M}}d^{3}x\hspace{.05cm}\epsilon^{\lambda\mu\nu} \bigg\{
-\sigma e_{\lambda}^{a}d_{[\mu}\Omega_{\nu]a}-\frac{\sigma}{2}
e_{\lambda}^{a}\varepsilon_{abc}\Omega_{[\mu}^{b}\Omega_{\nu]}^{c}
+\frac{\Lambda_{0}}{6}e_{\lambda}^{a}\varepsilon_{abc}e_{[\mu}^{b}e_{\nu]}^{c}
+h_{\lambda}^{a}d_{[\mu}e_{\nu]a}+\cr\nonumber\\[-1.5mm]
&&\hspace{1.5cm}+h_{\lambda}^{a}\varepsilon_{abc}
\Omega_{[\mu}^{b}e_{\nu]}^{c}+\frac{1}{2\mu}\Omega_{\lambda}^{a}d_{[\mu}\Omega_{\nu] a} 
+\frac{1}{6\mu}\Omega_{\lambda}^{a}
\varepsilon_{abc}\Omega_{[\mu}^{b}\Omega_{\nu]}^{c}+\sigma\alpha 
e_{\lambda}^{a}d_{[\mu}h_{\nu]a}
+\sigma\alpha e_{\lambda}^{a}\varepsilon_{abc}\Omega_{[\mu}^{b}
h_{\nu]}^{c}-\cr\nonumber\\[-1.5mm]
&&\hspace{1.5cm}-\frac{\alpha}{2}
(1+\sigma\alpha)e_{\lambda}^{a}\varepsilon_{abc}h_{[\mu}^{b}h_{\nu]}^{c}-\frac{\alpha}{2\mu}\Omega_{\lambda}^{a}d_{[\mu}h_{\nu]a}
-\frac{\alpha}{2\mu}\Omega_{\lambda}^{a}
\varepsilon_{abc}\Omega_{[\mu}^{b}h_{\nu]}^{c}-\frac{\alpha}{2\mu}
h_{\lambda}^{a}d_{[\mu}\Omega_{\nu]a}-
\cr\nonumber\\[-1.5mm]
&&\hspace{1.5cm}-\frac{\alpha}{4\mu}h_{\lambda}^{a}\varepsilon_{abc}\Omega_{[\mu}^{b}
\Omega_{\nu]}^{c}+\frac{\alpha^{2}}{2\mu}
h_{\lambda}^{a}d_{[\mu}h_{\nu]a}+\frac{\alpha^{2}}{2\mu}h_{\lambda}^{a}\varepsilon_{abc}
\Omega_{[\mu}^{b}h_{\nu]}^{c}-
\frac{\alpha^{3}}{6\mu}h_{\lambda}^{a}\varepsilon_{abc}
h_{[\mu}^{b}h_{\nu]}^{c}\bigg\},
\eea
where in our notation, $"A_{[\mu}B_{\nu]} = A_{\mu}B_{\nu} 
-A_{\nu}B_{\mu}"$ and 
$"\epsilon^{\lambda\mu\nu}"$ is $"\pm1"$\footnote{In our notation 
$\epsilon^
{\lambda\mu\nu}=-e\tilde{\epsilon}^{\lambda\mu\nu}$ with $\tilde
{\epsilon}^{\lambda\mu\nu}=\frac{1}{\sqrt{-g}}$.}. In our notation, from the first three terms
in the above action we have
\be
S = \int_{\mathcal{M}}d^{3}x\hspace{.05cm} \epsilon^{\lambda\mu\nu}\bigg\{-\sigma e_{\lambda}^{a}(\partial_{\mu}\Omega_{\nu a}-\partial_{\nu}\Omega_{\mu a})
-\sigma\varepsilon_{abc}e_{\lambda}^{a}\Omega_{\mu}^{b}\Omega_{\nu}^{c}+\frac{\Lambda_{0}}{3}
\varepsilon_{abc}e_{\lambda}^{a}e_{\mu}^{b}e_{\nu}^{c}\bigg\},
\ee
which is the well-known action of three dimensional gravity\cite{Witten:1988hc}.
%%%%%%%%%%%%%%%%%%%%%%%%%%%%%%%%%
\section{Divergent terms of on-shell action}\label{Div}
In this appendix we present explicit forms of divergent terms appearing in the on-shell action.
%with respect to parameters of the theory $\sigma,\Lambda_{0},\mu,\alpha$. 
Indeed, by making use of  (\ref{As.Critical.MMG}) the divergent terms  
of the on-shell action (\ref{Div.OnShell}) are found\footnote{
%{To be more precise, the equations in
%last line of (\ref{As.Critical.MMG}) not be used to find the divergent form 
%of on-shell action. That equations are relevant for studying the Ward identity which is related  
%to the divergence of stress tensor.}\raisebox{2mm}{,}\footnote{
To simplify these expressions we have assumed  that 
$\sigma^{2} =1$.  Moreover, throughout this paper we have also assumed that
the $\chi^{(0)}_{u}$ is sufficiently  small. Maple codes containing the calculations may be 
provided by the authors upon request.}
{\setlength\arraycolsep{2pt}
\bea
&&\mathcal{S}^{(0)}_{1/r} = \frac{1}{2}\mu^{2}C^{3}\alpha^{3}
+\frac{3}{2}\sigma\mu^{2}C^{2}\alpha^{2}+\frac{1}{4}
(-3C+6C^{2}\mu^{2})\alpha+\frac{1}{2}\sigma-\frac{1}{2}\Lambda_{0},\cr\nonumber\\[-2.5mm]
&&\mathcal{S}^{(1)}_{\log^{2}(r)} = LLS~\pv^{2}\chi^{(0)}_{u},\cr\nonumber\\[-2mm]
&&\mathcal{S}^{(1)}_{\log(r)} = \frac{1}{4\mu^{2}(1+\sigma\alpha)^{2}}
\bigg\{LL^{(1)}\pv^{2}\chi^{(0)}_{u}+2LL^{(2)}\big(\pu^{2}\varphi^{(0)}_{v}
+\pv^{2}\varphi^{(0)}_{u}-2\pu\pv \tilde{\varphi}^{(0)}_{v}\big)\bigg\},\cr\nonumber\\[-2mm]
&&\mathcal{S}^{(1)}_{1/r} = LR~\tilde{\varphi}^{(0)}_{v},\cr\nonumber\\[-2.5mm]
&&\mathcal{S}^{(2)}_{\log^{3}(r)} =  QLT~
\chi^{(0)}_{u}\chi^{(1)}_{v},\cr\nonumber\\[-2mm]
&&\mathcal{S}^{(2)}_{\log^{2}(r)}= \frac{1}{8\mu(1+\sigma\alpha)^{2}}\bigg\{
QLS^{(1)}(\chi^{(0)}_{u}\varphi^{(2)}_{v}+\varphi^{(0)}_{u}\chi^{(1)}_{v})
+QLS^{(2)}\chi^{(0)}_{u}\pv^{2}\tilde{\varphi}^{(0)}_{v}-\cr
&&\hspace{4.25cm}-\frac{1}{2}QLS^{(3)}\chi^{(0)}_{u}\pu\pv \varphi^{(0)}_{v}\bigg\},\cr\nonumber\\[-2.5mm]
&&\mathcal{S}^{(2)}_{\log(r)/r} =  QLR~\chi^{(0)}_{u}\varphi^{(0)}_{v},\cr\nonumber\\[-2mm]
&&\mathcal{S}^{(2)}_{\log(r)}= \frac{1}{(1+\sigma\alpha)^{4}} 
\bigg\{-\frac{1}{\mu^{3}} QL^{(1)}
\chi^{(0)}_{u}\chi^{(1)}_{v}+\frac{1}{2\mu^{2}} QL^{(2)}(-\chi^{(0)}_{u}\varphi^{(1)}_{v}
+\varphi^{(0)}_{u}\chi^{(1)}_{v})+\cr\nonumber\\[-.75mm]
&&\hspace{3.75cm}+\frac{1}{4\mu}QL^{(3)}\varphi^{(0)}_{u}\varphi^{(1)}_{v}+\frac{1}{4\mu}
QL^{(4)}\varphi^{(0)}_{v}\varphi^{(1)}_{u} 
+\frac{1}{4\mu^{2}} QL^{(5)}
\chi^{(0)}_{u}\pv^{2}\tilde{\varphi}^{(0)}_{v}+\cr\nonumber\\[-2mm]
&&\hspace{.8cm}+\frac{1}{4\mu^{2}}
QL^{(6)}\chi^{(0)}_{u}\pu\pv \varphi^{(0)}_{v}+\frac{1}{4}QL^{(7)}
(\tilde{\varphi}^{(0)}_{v}\pv^{2}\varphi^{(0)}_{u}
+\tilde{\varphi}^{(0)}_{v}\pu^{2}\varphi^{(0)}_{v}
-\frac{1}{2}\tilde{\varphi}^{(0)}_{v}\pu\pv \tilde{\varphi}^{(0)}_{v})\bigg\},\cr\nonumber\\[-2mm]
&&\mathcal{S}^{(2)}_{1/r} = \frac{1}{(1+\sigma\alpha)^{2}}
\bigg\{-\frac{1}{4\mu^{2}}QR^{(1)}\chi^{(0)}_{u}\varphi^{(0)}_{v}
-\frac{1}{4}QR^{(2)}
(\varphi^{(0)}_{u}\varphi^{(0)}_{v}-\tilde{\varphi}^{(0) 2}_{v})\bigg\},
\eea}%
where
\bea
&&LLS = \frac{1}{4}\mu^{2}C^{3}\alpha^{3}
+\frac{3}{4}\sigma\mu^{2}C^{2}\alpha^{2}+(-\frac{3}{8}C
+\frac{3}{4}\mu^{2}C^{2})\alpha+\frac{1}{4\mu}-\frac{3}{4}\sigma
-\frac{1}{4}\Lambda_{0}, \cr\nonumber\\[-1mm]
&&LL^{(1)} =2C^{3}\mu^{4}\alpha^{5}+(6C^{2}\mu^{4}\sigma +4C^{3}\mu^{4}
\sigma)\alpha^{4}+(18C^{2}\mu^{4}+2C^{3}\mu^{4}-3C\mu^{2})\alpha^{3}+\cr
&&\hspace{1.1cm}+(-2\Lambda_{0}\mu^{2}-6\sigma 
C\mu^{2}+18\sigma C^{2}\mu^{4}-6\sigma\mu^{2})\alpha^{2}
+(6C^{2}\mu^{4}-12\mu^{2} -3C\mu^{2}-\cr
&&\hspace{1.1cm}-4\sigma\Lambda_{0}\mu^{2}-4)\alpha-6\sigma\mu^{2}-2\Lambda_{0}
\mu^{2},\cr\nonumber\\[-2mm]
&&LL^{(2)} =C^{3}\mu^{4}\alpha^{5}
+2\sigma(C^{3}+\frac{3}{2}C^{2})\mu^{4}\alpha^{4}+\mu^{2}(-\frac{3}{2}C
+C^{3}\mu^{2}+9C^{2}\mu^{2})\alpha^{3}+\cr
&&\hspace{1.1cm}+\mu(1+9\sigma C^{2}
\mu^{3}-3\sigma\mu-\Lambda_{0}\mu -3\sigma C\mu)\alpha^{2}+
\mu(3C^{2}\mu^{3}-6\mu -\frac{3}{2}C\mu-\cr
&&\hspace{1.1cm}-2\sigma\Lambda_{0}\mu +2\sigma)\alpha +\mu -3\sigma\mu^{2}
-\Lambda_{0}\mu^{2},\cr\nonumber\\[-2mm]
&&LR = C^{3}\mu^{2}\alpha^{3}
+3\sigma C^{2}\mu^{2}\alpha^{2}+
(-\frac{3}{2}C+3C^{2}\mu^{2})\alpha+\sigma -\Lambda_{0},
\eea
and
\bea
&&QLT = \frac{1}{6}C^{3}\mu^{2}\alpha^{3}
+(\frac{1}{2}\sigma C^{2}\mu^{2}+\frac{1}{6}C^{2}\mu )\alpha^{2}
+(\frac{1}{2}C^{2}\mu^{2}+\frac{1}{3}\sigma C\mu -\frac{1}{4}C)\alpha-\cr
&&\hspace{1cm}-\frac{1}{6}\sigma-\frac{1}{6}\Lambda_{0}-\frac{1}{6\mu}
+\frac{1}{3}C\mu ,\cr\nonumber\\[-1mm]
&&QLS^{(1)} = 2C^{3}\mu^{3}\alpha^{5}+(4\sigma C^{3}\mu^{3}+
2C^{2}\mu^{2}+6\sigma C^{2}\mu^{3})\alpha^{4}+(-3C+18 C^{2}\mu^{3}+\cr
&&\hspace{1.4cm}+4\sigma C\mu^{2}\mu +4\sigma\mu^{2} C^{2}+
2\mu^{3} C^{3})\alpha^{3}+(-2+12C\mu^{2}-2\sigma\mu +18\sigma\mu^{3} 
C^{2}+\cr
&&\hspace{1.4cm}+2C^{2}\mu^{2}-6\sigma\mu C-2\mu\Lambda_{0})\alpha^{2}
+(12\sigma C\mu^{2}-4\sigma-4\mu-3 C\mu- 4\sigma\mu\Lambda_{0}+\cr
&&\hspace{1.4cm}+6\mu^{3}C^{2})\alpha -2-2\sigma\mu- 2\mu\Lambda_{0}
+4C\mu^{2},\cr\nonumber\\[-2mm]
&&QLS^{(2)}=2C^{3}\mu^{3}\alpha^{5}+(4\sigma C^{3}\mu^{3}
+6\sigma C^{2}\mu^{3})\alpha^{4}+(2\mu^{3}C^{3}-C\mu +18C^{2}\mu^{3})\alpha^{3}+\cr
&&\hspace{1.4cm}+(18\sigma C^{2}\mu^{3}-2\sigma\mu 
-2\Lambda_{0}\mu-2\sigma C\mu)\alpha^{2}+(-4\mu -C\mu -4\sigma\mu\Lambda_{0}
+6C^{2}\mu^{3})\alpha -\cr
&&\hspace{1.4cm}-2\mu\Lambda_{0}-2\sigma\mu,\cr\nonumber\\[-2mm]
&&QLS^{(3)}=2C^{3}\mu^{3}\alpha^{5}+(4\sigma\mu^{3} C^{3}
-2C^{2}\mu^{2}+6\sigma\mu^{3}C^{2})\alpha^{4}+(C\mu+ 18\mu^{3}C^{2}-4\sigma C\mu^{2}-\cr
&&\hspace{1.4cm}-4\sigma\mu^{2} C^{2}+2\mu^{3} C^{3})
\alpha^{3}+(2-12C\mu^{2}-2\sigma\mu+ 18\sigma C^{2}\mu^{3}-2C^{2}\mu^{2}+
2\sigma\mu C-\cr
&&\hspace{1.4cm}-2\mu\Lambda_{0})\alpha^{2}
+(-12\sigma C\mu^{2}+4\sigma-4\mu +C\mu -4
\sigma\mu\Lambda_{0}+6\mu^{3}C^{2})
\alpha +2-2\sigma\mu -\cr
&&\hspace{1.4cm}- 2\mu\Lambda_{0}-4C\mu^{2},\cr\nonumber\\[-2mm]
&&QLR = -\frac{1}{2}\mu^{2}C^{3}\alpha^{3}
-\frac{3}{2}\sigma\mu^{2}C^{2}\alpha^{2}-\frac{1}{2}(
-\frac{3}{2}C+3C^{2}\mu^{2})\alpha-\frac{1}{2}\sigma+\frac{1}{2}
\Lambda_{0},\cr\nonumber\\[-2mm]
&&QL^{(1)} = C\mu^{3}\alpha^{5}
+(-2\sigma\mu^{3}-4\mu^{2}+4C\mu^{2}+4\sigma C\mu^{3})
\alpha^{4}+(8\sigma C\mu^{2}-8\sigma\mu^{2}-16\mu +\cr
&&\hspace{1.12cm}+ 8C\mu +6C\mu^{3}-8\mu^{3})\alpha^{3}
+(4\sigma C\mu^{3}+4C\mu^{2}-8\mu^{2}
-12\sigma\mu^{3}-8-24\sigma\mu)\alpha^{2}+\cr
&&\hspace{1.12cm}+(-8\sigma\mu^{2}-8\mu^{3}-8\mu+
C\mu^{3})\alpha -2\sigma\mu^{3} -4\mu^{2},\cr \nonumber\\[-2mm]
&&
QL^{(2)} = C^{2}\mu^{3}\alpha^{6}
+(4\sigma\mu^{3}C^{2}-2C\mu^{2}+4C^{2}\mu^{2}+2\sigma 
C\mu^{3})\alpha^{5}+(10C\mu^{3}-2\sigma\mu^{2}+\cr
&&\hspace{1.12cm}+6C^{2}\mu^{3}-3\mu +8\sigma\mu^{2} C^{2})\alpha^{4}+
(12C\mu^{2}+20\sigma C\mu^{3}-12\sigma\mu- 8\mu^{2}-6+4C^{2}\mu^{2}+\cr
&&\hspace{1.12cm}+4\sigma C^{2}\mu^{3})
\alpha^{3}+(C^{2}\mu^{3}+16\sigma C\mu^{2}-12\sigma\mu^{2}
+20C\mu^{3}-18\mu- 12\sigma)\alpha^{2}+\cr
&&\hspace{1.12cm}+(-8\mu^{2}-6+10\sigma C\mu^{3}+6C\mu^{2}-12\sigma\mu)\alpha
+2C\mu^{3}-3\mu- 2\sigma\mu^{2},\cr \nonumber\\[-2mm]
&&
QL^{(3)} =2C^{3}\mu^{3}\alpha^{7}
+(2C^{2}\mu^{2}+8\sigma C^{3}\mu^{3}+6\sigma C^{2}\mu^{3})
\alpha^{6}+(30C^{2}\mu^{3}+12C^{3}\mu^{3}-3C\mu +\cr
&&\hspace{1.12cm}+8\sigma C^{2}
\mu^{2}+4\sigma C\mu^{2})\alpha^{5}+(8\sigma C^{3}\mu^{3}+20 C\mu^{2}
+12 C^{2}\mu^{2}-12\sigma C \mu+ 60\sigma C^{2}\mu^{3}-\cr
&&\hspace{1.12cm}-2\sigma\mu- 2\mu \Lambda_{0}-2)\alpha^{4}+(-8\mu- 
8\sigma +2C^{3}\mu^{3}-18C \mu +40\sigma
C \mu^{2}+60 C^{2}\mu^{3}-\cr
&&\hspace{1.12cm}-8\sigma\mu\Lambda_{0}
+8\sigma C^{2}\mu^{2})\alpha^{3}+(-12-12\mu\Lambda_{0}+2C^{2}\mu^{2}-12\sigma\mu C
+40C\mu^{2}+30\sigma C^{2}\mu^{3}-\cr
&&\hspace{1.12cm}
-12\sigma\mu)\alpha^{2}+(-8\mu- 8\sigma+ 20\sigma C\mu^{2}
-3C\mu+ 6C^{2}\mu^{3}-8\sigma\mu\Lambda_{0})\alpha -2-2\sigma\mu -\cr
&&\hspace{1.12cm}- 2\mu\Lambda_{0}
+4C\mu^{2},\cr\nonumber\\[-2mm]
&&
QL^{(4)} =2C^{3}\mu^{3}\alpha^{7}+(-2C^{2}\mu^{2}+8\sigma C^{3}\mu^{3}+6\sigma C^{2}\mu^{3})\alpha^{6}+(30C^{2}\mu^{3}
+12C^{3}\mu^{3}+C\mu-\cr
&&\hspace{1.12cm}- 8\sigma C^{2}\mu^{2}-4\sigma 
C\mu^{2})\alpha^{5}+(8\sigma C^{3}\mu^{3}-20 C\mu^{2}-12 C^{2}\mu^{2}
+4\sigma C \mu + 60\sigma C^{2}\mu^{3}-\cr
&&\hspace{1.12cm}-2\sigma\mu - 2\mu \Lambda_{0}+2)\alpha^{4}+(-8\mu + 8\sigma 
+2C^{3}\mu^{3}+6C \mu -40\sigma C \mu^{2}+60 C^{2}\mu^{3}-\cr
 &&\hspace{1.12cm}-8\sigma\mu\Lambda_{0}-8\sigma C^{2}\mu^{2})
\alpha^{3}+(12-12\mu\Lambda_{0}-2C^{2}\mu^{2}+4\sigma\mu C-
40C\mu^{2}+30\sigma C^{2}\mu^{3}-\cr
&&\hspace{1.12cm}-12\sigma\mu)\alpha^{2}+(-8\mu+ 8\sigma - 20\sigma C\mu^{2}
+C\mu+ 6C^{2}\mu^{3}-8\sigma\mu\Lambda_{0})\alpha
+2-2\sigma\mu -\cr
&&\hspace{1.12cm}-2\mu\Lambda_{0}-4C\mu^{2},\cr\nonumber\\[-2mm]
&&
QL^{(5)} = 2C^{3}\mu^{4}\alpha^{7}+(6\sigma C^{2}
\mu^{4}+8\sigma C^{3}\mu^{4})\alpha^{6}+(12C^{3}\mu^{4}
+8C^{2}\mu^{2}+30C^{2}\mu^{4}-3C\mu^{2})\alpha^{5}+\cr
&&\hspace{1.12cm}+(16\sigma C^{2}\mu^{2}
+4\sigma C \mu^{2}+8\sigma C^{3}\mu^{4}-2\mu^{2}\Lambda_{0}
-4\mu-6\sigma\mu^{2}+60\sigma C^{2}\mu^{4})\alpha^{4}+\cr
&&\hspace{1.12cm}+(-20+30C \mu^{2}-16\sigma\mu+ 2C^{3}\mu^{4}-24\mu^{2}
+60 C^{2}\mu^{4}-8\sigma\mu^{2}\Lambda_{0}+8C^{2}\mu^{2})
\alpha^{3}+\cr
&&\hspace{1.12cm}+(-24\mu- 40\sigma +30\sigma C^{2}\mu^{4}-12 \mu^{2}\Lambda_{0}
+36\sigma C \mu^{2}-36\sigma\mu^{2})\alpha^{2}+(-16\sigma\mu -20-\cr
&&\hspace{1.12cm}-24\mu^{2}+6C^{2}\mu^{4}+13C\mu^{2}-8\sigma\mu^{2}\Lambda_{0})
\alpha-4\mu- 2\mu^{2}\Lambda_{0}-6\sigma\mu^{2},\cr\nonumber\\[-2mm]
&&
QL^{(6)} = C^{2}\mu^{3}\alpha^{6}+(-4C^{2}\mu^{2}+4\sigma C^{2}\mu^{3}
+2\sigma C\mu^{3})\alpha^{5}+(2\sigma\mu^{2}-8\sigma C^{2}\mu^{2}
+\mu +6 C^{2}\mu^{3}-\cr
&&\hspace{1.12cm}-8\sigma C\mu^{2}+10C \mu^{3})
\alpha^{4}+(8\mu^{2}+6+4\sigma\mu- 4C^{2}\mu^{2}
+4\sigma C^{2}\mu^{3}+20\sigma C\mu^{3}-\cr
&&\hspace{1.12cm}-24 C\mu^{2})\alpha^{3}+
(C^{2}\mu^{3}+20 C\mu^{3}+12\sigma\mu^{2}+12\sigma - 
24\sigma C\mu^{2}+6\mu)\alpha^{2}+(10\sigma C\mu^{3}+\cr
&&\hspace{1.12cm}+6-8C\mu^{2}+8\mu^{2}+4\sigma\mu)\alpha
+2\sigma\mu^{2}+\mu+ 2C\mu^{3}, \cr \nonumber\\[-2mm]
&&
QL^{(7)} = 2C^{3}\mu^{2}\alpha^{7} +(8\sigma C^{3}\mu^{2}
+6\sigma C^{2}\mu^{2})\alpha^{6}+(-C+30C^{2}\mu^{2}+12C^{3}\mu^{2})\alpha^{5}
+(8\sigma C^{3}\mu^{2}-\cr
&&\hspace{1.12cm}-2\Lambda_{0}+60\sigma C^{2}\mu^{2}-
4\sigma C-2\sigma)\alpha^{4}+(-6C+2C^{3}\mu^{2}+60 C^{2}\mu^{2}-8\sigma\Lambda_{0}
-8)\alpha^{3}+\cr
&&\hspace{1.12cm}+(-4\sigma C-12\sigma-12\Lambda_{0}+30\sigma
C^{2}\mu^{2})\alpha^{2}+(-8-8\sigma\Lambda_{0}-C+6C^{2}\mu^{2})\alpha -2\Lambda_{0} -\cr
&&\hspace{1.12cm}-2\sigma,\cr\nonumber\\[-2mm]
&&QR^{(1)}=2C^{3}\mu^{4}\alpha^{5}
+(4\sigma C^{3}\mu^{4}+6\sigma C^{2}\mu^{4}-2C^{2}\mu^{3})
\alpha^{4}+(2C^{3}\mu^{4}-8C^{2}\mu^{2}-4\sigma C^{2}\mu^{3}-\cr
&&\hspace{1.12cm}-4\sigma C \mu^{3}+18C^{2}\mu^{4}+5C\mu^{2})\alpha^{3}
+(8C\mu- 2C^{2}\mu^{3}+18\sigma C^{2}\mu^{4}-12C \mu^{3}
-2\mu^{2}\Lambda_{0}-\cr
&&\hspace{1.12cm}-2\mu- 6\sigma C\mu^{2}-2\sigma\mu^{2})\alpha^{2}
+(-4+12\sigma\mu- 4\mu^{2}-11C\mu^{2}-4\sigma\mu^{2}\Lambda_{0}
-12\sigma C\mu^{3} +\cr
&&\hspace{1.12cm}+6C^{2}\mu^{4})\alpha- 2\mu- 2\sigma\mu^{2}-4C\mu^{3}
-2\mu^{2}\Lambda_{0},\cr\nonumber\\[-2mm]
&&QR^{(2)}=2C^{3}\mu^{2}\alpha^{5}
+(6\sigma C^{2}\mu^{2}+4\sigma C^{3}\mu^{2})\alpha^{4}
+(2C^{3}\mu^{2}-3C+18C^{2}\mu^{2})\alpha^{3}+(18\sigma C^{2}\mu^{2}-\cr
&&\hspace{1.12cm}-6\sigma C- 2\Lambda_{0}+2\sigma)\alpha^{2}
+(-3C+4-4\sigma\Lambda_{0}+6C^{2}\mu^{2})\alpha + 2\sigma- 2\Lambda_{0}.
\eea
%%%%%%%%%%%%%%%%%%%%%%%%%%%%%%%%%%%%%%%%%%%
\section{Coefficients in Counterterm Action}\label{EVC}
In this appendix, explicit values of the coefficients appearing in the action of  counterterms 
for two cases $\sigma =-1$ and $\sigma =1$ are presented.
\vspace{-.25cm} 
\subsection{$\sigma =-1$}
\vspace{-.75cm}
\bea
&&a_{1}=-\frac{15\mu^{3}-87\mu^{2}+16\mu +60}{32\mu^{2}},~~~~
a_{2}=a_{4}=\frac{5\mu -9}{16\mu},~~~~~a_{3}=-\frac{1}{\mu},\cr\nonumber\\[-2mm]
&&a_{5} =\frac{\mu^{3}-9\mu^{2}+48\mu -60}{32\mu^{2}},~~~~~
~~~a_{6}=\frac{(5\mu -1)(\mu -2)^{2}}{16\mu^{2}},~~~~~~
a_{7}=-\frac{3\mu-1}{64\mu},\cr\nonumber\\[-2mm]
&&b_{1} =-\frac{9\mu^{3}
-47\mu^{2}+88\mu -52}{16\mu^{2}},~~~
b_{2} = \frac{3\mu^{3}
-16\mu^{2}+34\mu -22}{8\mu^{2}},\cr\nonumber\\[-2mm]
&&b_{3}=-\frac{1}{2}b_{4}=
\frac{1}{4}b_{8}=4c_{5} =\frac{\mu -1}{8\mu},~~~
b_{5}=-\frac{4\mu^{3}-21\mu^{2}+40\mu -24}{8\mu^{2}},\cr\nonumber\\[-2mm]
&&b_{6}=\frac{4\mu^{3}-21\mu^{2}+38\mu -22}{8\mu^{2}},~~~~
b_{7}=\frac{(\mu-2)^{2}(9\mu^{3}-41\mu^{2}
+76\mu -48)}{32\mu^{3}},\cr
&&c_{1}=-4c_{2}=-\frac{4}{3}c_{3}=\frac{(\mu -1)(\mu -2)^{2}}{16\mu^{2}},~~
c_{4}=6d_{1} = -\frac{(\mu -1)(\mu -2)^{4}}{64\mu^{3}}.
\eea
\subsection{$\sigma =1$}
\vspace{-.75cm}
\bea
&&a_{1} = \frac{5\mu^{3}+45\mu^{2}-32\mu -68}{32\mu^{2}},~~~~~
a_{2}=a_{5} = -\frac{5\mu +9}{16\mu},~~~~a_{3} = -\frac{(5\mu +1)
(\mu +2)^{2}}{16\mu^{2}},\cr\nonumber\\[-2mm]
&&a_{4}=-\frac{1}{\mu},~~~~~
a_{6} = -\frac{11\mu^{3}+51\mu^{2}+96\mu +68}{32\mu^{2}},~~~~~~
a_{7} = -\frac{3\mu^{3}+15\mu^{2}+44\mu +40}{128\mu(\mu+2)^{2}},\cr\nonumber\\[-2mm]
&&b_{1} = -\frac{9\mu^{3}+29\mu^{2} +52\mu +40}{32\mu^{2}},~~~~
b_{2} = -\frac{(\mu +1)(3\mu^{2} +20\mu +40)}{32\mu^{2}},\cr\nonumber\\[-2mm]
&&b_{3}=-\frac{1}{2}b_{4}=\frac{1}{4}b_{7}=4c_{5}
 = -\frac{\mu +1}{8\mu},~~~~
b_{5} = \frac{(\mu +1)(7\mu^{2}
+36\mu +48)}{32\mu^{2}},\cr\nonumber\\[-2mm]
&&b_{6} = \frac{11\mu^{3} +39\mu^{2}
+76\mu +56}{32\mu^{2}},~~~~b_{8}= -\frac{9\mu^{3} +41\mu^{2}
+76\mu +48}{16\mu(\mu +2)^{2}},\cr
&&c_{1}=-4c_{2}=-\frac{4}{3}c_{3} = -\frac{(\mu +1)(\mu +2)^{2}}{16\mu^{2}}
,~~~c_{4}=6d_{1} = \frac{(\mu +1)(\mu +2)^{4}}{64\mu^{3}}.
\eea
%%%%%%%%%%%%%%%%%%%%%%%%%%%%%%%%%%%%%%
\section{Local Finite Terms in On-Shell Action}\label{LFO}
In this appendix , explicit forms of the local terms in the renormalized on-shell
action for two cases $\sigma =-1$ and $\sigma =1$ are provided. 
\vspace{-.25cm} 
\subsection{$\sigma =-1$}
\vspace{-.75cm} 
\bea
&&S_{ren}\big{|}_{local}= \int d^{2}x \bigg\{\frac{2(\mu^{3}-4\mu^{2}+4)}
{\mu(\mu -2)^{2}}\chi^{(0)}_{u}\pu\pv
\varphi^{(0)}_{v} -\frac{29\mu^{3}-141\mu^{2}+240\mu 
-140}{4\mu(\mu -2)^{2}}\chi^{(0)}_{u}\pv^{2}\tilde{\varphi}^{(0)}_{v}\cr\nonumber\\[-1mm]
&&-\frac{15\mu -19}{4\mu}\tilde{\varphi}^{(0)}_{v}\pv^{2}\varphi^{(0)}_{u} 
-\frac{15\mu -19}{4\mu}\tilde{\varphi}^{(0)}_{v}\pu^{2}\varphi^{(0)}_{v}
+\frac{19\mu -27}{4\mu}\tilde{\varphi}^{(0)}_{v}\pu\pv \tilde{\varphi}^{(0)}_{v}
-\frac{\mu +7}{4\mu}\varphi^{(0)}_{v}\pu\pv \varphi^{(0)}_{u}\bigg\}.\nn \\
\eea
\vspace{-1.75cm} 
\subsection{$\sigma =1$}
\vspace{-.75cm} 
\bea
&&S_{ren}\big{|}_{local}= \int d^{2}x \bigg\{
-\frac{2(\mu^{3}+4\mu^{2}-4)}{\mu(\mu +2)^{2}}\chi^{(0)}_{u}\pu\pv
\varphi^{(0)}_{v} +\frac{39\mu^{3}+183\mu^{2}+288\mu 
+148}{4\mu(\mu +2)^{2}}\chi^{(0)}_{u}\pv^{2}\tilde{\varphi}^{(0)}_{v}\cr\nonumber\\[-1mm]
&&+\frac{15\mu +19}{4\mu}\tilde{\varphi}^{(0)}_{v}\pv^{2}\varphi^{(0)}_{u} 
+\frac{15\mu +19}{4\mu}\tilde{\varphi}^{(0)}_{v}\pu^{2}\varphi^{(0)}_{v}
-\frac{7\mu^{3}+36\mu^{2}+65\mu +39}{\mu(\mu +2)^{2}}
\tilde{\varphi}^{(0)}_{v}\pu\pv \tilde{\varphi}^{(0)}_{v}\cr\nonumber\\[-1mm]
&&-\frac{2\mu^{3}+11\mu^{2}+25\mu +19}{\mu (\mu +2)^{2}}\varphi^{(0)}_{v}
\pu\pv \varphi^{(0)}_{u}\bigg\}.
\eea
%%%%%%%%%%%%%%%%%%%%%%%%%%%%%%%%
\vspace{-1cm} 
\section{Local Terms in One-Point Functions}\label{LTO}
In this appendix explicit forms of the local terms in one-point functions
are given. The contributions of these local terms to the two-point
functions are contact terms. It is important to note that these local terms  contribute
to the boundary charges such as mass, angular momentum and entropy
of black hole solutions. 
%Because studying these thermodynamic quantities at critical
%value $M=-1$ is not the subject of this paper 
%we decide to present the form of these local terms
%in this appendix.
\vspace{-.5cm} 
\subsection{$\sigma =-1$}
\vspace{-1.2cm} 
\bea
&&\langle T_{uu}\rangle |_{local} =\frac{1}{8G}\left(
\frac{2(\mu^{3}-4\mu^{2}+4)}{\mu(\mu -2)^{2}}
\pu\pv \chi^{(0)}_{u}-\frac{15\mu -19}{4\mu}\pu^{2}
\tilde{\varphi}^{(0)}_{v}-\frac{\mu +7}{4\mu}\pu\pv \varphi^{(0)}_{u}\right) \cr\nonumber\\[-1mm]
&&\langle T_{vv}\rangle |_{local} =\frac{1}{8G}\left(
-\frac{15\mu -19}{4\mu}\pv^{2}\tilde{\varphi}^{(0)}_{v}
-\frac{\mu +7}{4\mu}\pu\pv \varphi^{(0)}_{v}\right)\cr\nonumber\\[-1.5mm]
&&\langle T_{uv}\rangle |_{local} =\frac{1}{8G}\bigg(
-\frac{15\mu -19}{4\mu}\pv^{2}\varphi^{(0)}_{u}
-\frac{15\mu -19}{4\mu}\pu^{2}\varphi^{(0)}_{v}+\frac{19\mu -27}{2\mu}
\pu\pv \tilde{\varphi}^{(0)}_{v}\cr
&&\hspace{1.79cm}-\frac{29\mu^{3}-141\mu^{2}+240\mu 
-140}{4\mu(\mu -2)^{2}}\pv^{2}\chi^{(0)}_{u}\bigg),\cr\nonumber\\[-1.5mm]
&&\langle t_{vv}\rangle |_{local} =-\frac{1}{8G}
\left(\frac{2(\mu^{3}-4\mu^{2}+4)}{\mu(\mu -2)^{2}}
\pu\pv \varphi^{(0)}_{v}-\frac{29\mu^{3}-141\mu^{2}+240\mu 
-140}{4\mu(\mu -2)^{2}}\pv^{2} \tilde{\varphi}^{(0)}_{v}\right).\nn \\
\eea
\vspace{-1.75cm} 
\subsection{$\sigma =1$}
\vspace{-1.1cm} 
\bea
&&\langle T_{uu}\rangle |_{local} =\frac{1}{8G}\bigg(-
\frac{2(\mu^{3}+4\mu^{2}-4)}{\mu(\mu +2)^{2}}\pu\pv \chi^{(0)}_{u}
+\frac{15\mu +19}{4\mu}\pu^{2}
\tilde{\varphi}^{(0)}_{v}-\cr\nonumber\\[-1mm]
&&\hspace{1.79cm}-\frac{2\mu^{3}+11\mu^{2}+25\mu +19}
{\mu (\mu +2)^{2}}\pu\pv \varphi^{(0)}_{u}\bigg), \cr\nonumber\\[-1mm]
&&\langle T_{vv}\rangle |_{local} =\frac{1}{8G}\left(
\frac{15\mu +19}{4\mu}\pv^{2}\tilde{\varphi}^{(0)}_{v}
-\frac{2\mu^{3}+11\mu^{2}+25\mu +19}{\mu (\mu +2)^{2}}
\pu\pv \varphi^{(0)}_{v}\right),\cr\nonumber\\[-1mm]
&&\langle T_{uv}\rangle |_{local} =\frac{1}{8G}\bigg(\frac{15\mu 
+19}{4\mu}\pv^{2}\varphi^{(0)}_{u}
+\frac{15\mu +19}{4\mu}\pu^{2}\varphi^{(0)}_{v}-2
\frac{7\mu^{3}+36\mu^{2}+65\mu +39}{\mu(\mu +2)^{2}}
\pu\pv \tilde{\varphi}^{(0)}_{v}+\cr\nonumber\\[-1mm]
&&\hspace{1.79cm}+\frac{39\mu^{3}+183\mu^{2}+288\mu 
+148}{4\mu(\mu +2)^{2}}\pv^{2}\chi^{(0)}_{u}\bigg)\cr\nonumber\\[-1mm]
&&\langle t_{vv}\rangle |_{local} =-\frac{1}{8G}\left(-
\frac{2(\mu^{3}+4\mu^{2}-4)}{\mu(\mu +2)^{2}}\pu\pv \varphi^{(0)}_{v}
 +\frac{39\mu^{3}+183\mu^{2}+288\mu 
+148}{4\mu(\mu +2)^{2}}\pv^{2} \tilde{\varphi}^{(0)}_{v}\right).\nn \\
\eea

\end{document}